\documentclass[12pt]{article}
\usepackage{amsmath,amssymb,amsfonts,amsthm,amscd}
\usepackage[english]{babel}
\usepackage[toc]{appendix}
\usepackage{caption}
\usepackage[cp1251]{inputenc}
\newcommand{\be}{\begin{equation}}
\newcommand{\ee}{\end{equation}}
\newcommand{\bea}{\begin{eqnarray}}
\newcommand{\eea}{\end{eqnarray}}
\newcommand{\nn}{\nonumber \\}
\newcommand{\p}[1]{(\ref{#1})}
\newcommand{\lb}{\label}

\numberwithin{equation}{section}

\begin{document}
\begin{titlepage}
\begin{flushright}
JINR E2-2015-57
\end{flushright}
\vfill
\vfill
\begin{center}
\baselineskip=16pt {\Large\bf $SU(2|1)$ mechanics and harmonic superspace}

\vskip 0.3cm {\large {\sl }} \vskip 10.mm {\bf $\;$  E. Ivanov$^{\,a}$,  $\;$ S. Sidorov$^{\,b}$
}
\vspace{1cm}

{\it Bogoliubov Laboratory of Theoretical Physics, JINR, \\
141980 Dubna, Moscow Region, Russia\\
}
\end{center}
\vfill

\par
\begin{center}
{\bf Abstract}
\end{center}
We define the worldline harmonic $SU(2|1)$ superspace and its analytic subspace
as a deformation of the flat ${\cal N}=4, d=1$ harmonic superspace.
The harmonic superfield description of the  two mutually mirror off-shell ${\bf (4,4,0)}$ $SU(2|1)$ supermultiplets is developed and the
corresponding invariant actions are presented, as well as the relevant classical and quantum supercharges.  Whereas the $\sigma$-model
actions exist for both types of the ${\bf (4,4,0)}$ multiplet, the
invariant Wess-Zumino term can be defined only for one of them, thus demonstrating non-equivalence of these multiplets in the $SU(2|1)$ case
as opposed to the flat ${\cal N}=4, d=1$ supersymmetry. A superconformal subclass of general $SU(2|1)$ actions
invariant under the trigonometric-type realizations of the supergroup $D(2,1;\alpha)$ is singled out. The superconformal Wess-Zumino actions
possess an infinite-dimensional supersymmetry forming the centerless ${\cal N}=4$ super Virasoro algebra. We solve a few simple instructive
examples of the $SU(2|1)$ supersymmetric quantum mechanics of the ${\bf (4,4,0)}$ multiplets and reveal the $SU(2|1)$ representation contents
of the corresponding sets of the quantum states.

\vspace{2.5cm}

\begin{quote}
\vfill \vfill \vfill \vfill \vfill \hrule width 5.cm \vskip 2.mm
{\small
\noindent $^a$ eivanov@theor.jinr.ru\\
\noindent $^b$ sidorovstepan88@gmail.com
}
\end{quote}
\end{titlepage}

%%%%%%%%%%%%%%%%%%%%%%%%%%%%%%%%%%%%%%%%%%%%%%%%%%%%%%%%%%%%%%%%%%%%%%%%%%%%%%%%%
\section{Introduction}

Motivated by a recent interest in theories with a curved rigid supersymmetry (see, e.g., \cite{FS,DFS,SamSor}), in \cite{DSQM,SKO, ISTconf} we defined and studied a new type
of supersymmetric quantum mechanics (SQM) based on the worldline realizations of the supergroup $SU(2|1)$ and its central extension. It can be viewed as a deformation
of the standard flat ${\cal N}=4$ supersymmetric mechanics by a mass parameter $m$. We defined the worldline $SU(2|1)$ superfields, which
are carriers of the off-shell $SU(2|1)$ multiplets with the $d=1$ field contents ${\bf (1, 4, 3)}$ and ${\bf (2, 4, 2)}$, constructed the
invariant superfield and component actions for them and explored the classical and quantum properties of the corresponding $SU(2|1)$
invariant $d=1$ systems. A few interesting features of these deformed SQM models were revealed. In particular, the SQM models associated with the
multiplet ${\bf (1, 4, 3)}$ reproduce the ``weak supersymmetry'' systems studied in \cite{WS}. The invariant $\sigma$-model Lagrangians of the chiral multiplet
${\bf (2, 4, 2)}$ contain, along with the standard kinetic $\sigma$-model type terms for the physical bosonic fields, also the Wess-Zumino (WZ) type terms
of the first order in time derivative, as well as some induced scalar potential terms, all proportional to the deformation parameter $m$.  As distinct from
the case of flat ${\cal N}=4, d=1$ supersymmetry, there exist two different types of the chiral ${\bf (2, 4, 2)}$ multiplet in the $SU(2|1)$ case \cite{SKO},
with essentially different properties of the related SQM models. The Hilbert spaces of the quantum states in $SU(2|1)$ SQM models contain unequal numbers of
the fermionic and bosonic excited states due to the presence of the so called atypical representations of $SU(2|1)$ in the spectrum. This is a general feature of
the $SU(2|1)$ SQM systems firstly observed in the pioneer paper \cite{WS} for the ${\bf (1, 4, 3)}$ models. Recently,
we studied superconformal properties of the $SU(2|1)$ SQM models built on the multiplets ${\bf (2, 4, 2)}$ and ${\bf (1, 4, 3)}$ \cite{ISTconf}
and found that the formulation in the $SU(2|1)$ superspace automatically yields the trigonometric realizations of the most
general ${\cal N} =4, d=1$ superconformal group $D(2,1;\alpha)$ in the conformal subclass of such models. We established a simple
criterion under which  one or another $SU(2|1)$ invariant action possesses superconformal  $D(2,1;\alpha)$ symmetry.

As was argued in \cite{root} in the component approach and in \cite{DeldIv1} in the superfield language, the basic multiplet of
${\cal N}=4, d=1$ supersymmetry is the so called ``root'' multiplet ${\bf (4, 4, 0)}$. All other multiplets and the SQM models associated with them
can be obtained from the root multiplet and the associate SQM models by some well-defined procedures: either by a sort of Hamiltonian reduction with
respect to some isometries of the on-shell ${\bf (4, 4, 0)}$ Lagrangians, or by gauging these isometries in the off-shell superfield approach
\cite{DeldIv1,DeldIv2,DeldIv3}. The natural superfield description of the multiplet ${\bf (4, 4, 0)}$, as well as of another important multiplet
${\bf (3, 4, 1)}$, is achieved in the framework of the harmonic ${\cal N}=4, d=1$ superspace \cite{IvLe}.
In particular, the $d=1$ harmonic superspace approach allows one to establish relations between various ${\cal N}=4, d=1$ multiplets
in the  manifestly off-shell supersymmetric  way and to understand the general target geometry
of the ${\bf (4, 4, 0)}$ superfield models \cite{DeldIv4}.

In view of the crucial role of the harmonic superspace approach \cite{HSS0,HSS} in the flat ${\cal N}=4$ SQM models it is natural
to expect that it could  be of equal importance for $SU(2|1)$ SQM models as deformations of the ${\cal N}=4$ ones.
The existence of the worldline  harmonic coset in the supergroup $SU(2|1)$ was observed in \cite{DSQM}
\footnote{Some other kind of harmonic extensions of $SU(2|1)$ was considered in \cite{KuSor}.}.
The main purpose of the present paper is to work out, in full generality,
the $d=1$ harmonic superspace formalism for  $SU(2|1)$, to describe
the deformed analogs of the flat  ${\bf (4, 4, 0)}$ multiplet within
this framework and to construct the relevant SQM models, in both the superfield and the component forms.

In section 2, we define the harmonic $SU(2|1)$ superspace $\zeta_H$ and its analytic subspace $\zeta_{(A)}$
as the proper supercosets of the harmonic extension of $SU(2|1)$, following the general construction of \cite{HSS}.
We give the explicit expressions for the relevant
covariant derivatives, both Grassmann and harmonic,  in the basis $\zeta_H$ and define the $SU(2|1)$ analyticity conditions.
In sections 3 and 4, we consider the $SU(2|1)$ harmonic superfields $q^{\pm a}$ and $Y^A, \bar{Y}^A$ describing
two types of the ${\bf (4,4,0)}$ supermultiplets which can be defined in the $SU(2|1)$ case.
The general $\sigma$-model type terms for both multiplets are constructed and it is shown that an external $SU(2|1)$ invariant WZ term
can be defined only for the mirror multiplet ${\bf (4,4,0)}$, in a crucial distinction from the flat ${\cal N}=4, d=1$ case.
We present general expressions for the relevant supercharges, in the classical and quantum cases, and explicitly
find the spectrum of the corresponding Hamiltonian for a few simple models. In section 5, we single out the superconformal
subclass of the $SU(2|1)$ SQM models for both types of the ${\bf (4,4,0)}$ multiplet  and show that they
exhibit the trigonometric realization of the superconformal group $D(2,1;\alpha)$, like the SQM models  considered
in \cite{HT,ISTconf}. The superconformal WZ terms reveal an infinite-dimensional superconformal symmetry corresponding to some centerless ${\cal N}=4$
super Virasoro algebra. In section 6, we summarize the results and outline some further directions of study. Some technical details are collected in Appendices
A, B, C.
\section{$SU(2|1)$ harmonic superspace}
\subsection{Superalgebra}
We start from the standard form of the superalgebra $su(2|1)$:
\bea
    &&\lbrace Q^{i}, \bar{Q}_{j}\rbrace = 2m I^i_j +2\delta^i_j \tilde H,\qquad
    \left[I^i_j,  I^k_l\right]
    = \delta^k_j I^i_l - \delta^i_l I^k_j\,,\nn
    &&\left[I^i_j, \bar{Q}_{l}\right] = \frac{1}{2}\,\delta^i_j\bar{Q}_{l}-\delta^i_l\bar{Q}_{j}\, ,\qquad \left[I^i_j, Q^{k}\right]
    = \delta^k_j Q^{i} - \frac{1}{2}\,\delta^i_j Q^{k},\nn
    &&\left[\tilde H, \bar{Q}_{l}\right]=\frac{m}{2}\,\bar{Q}_{l}\,,\qquad \left[\tilde H, Q^{k}\right]=-\frac{m}{2}\,Q^{k}. \label{algebra-1}
\eea
All other (anti)commutators are vanishing. It is a deformation of the standard ${\cal N}=4$, $d=1$ ``Poincar\'e'' superalgebra.
The bosonic subalgebra consists of the $SU(2)$ symmetry generators $I^i_j$ and the $U(1)$ symmetry generator $\tilde{H}$.
In the limit $m=0$, the generators $I^i_j$ become those of the $SU(2)$ automorphism group of
the standard ${\cal N}=4$, $d=1$ superalgebra, while $\tilde{H}$ turns into the $d=1$ time translation generator.

Actually, the flat ${\cal N}=4, d=1$ superalgebra has an additional automorphism group $SU^{\prime}(2)$ which rotates the $Q$ and $\bar{Q}$ generators
among each other\footnote{In what follows, the primed indices $i^{\prime},j^{\prime}$ will be associated with the $SU^{\prime}(2)$ doublets.}.
Upon deformation to the superalgebra \p{algebra-1}, one of these  $SU^{\prime}(2)$ generators, which we denote $F$,
still survives as an external $U(1)$ automorphism symmetry ($R$-symmetry) of the deformed superalgebra.
So one can extend \eqref{algebra-1} by the generator $F$ possessing non-zero commutation relations
with the supercharges only \cite{FS}:
\bea
    \left[F, \bar{Q}_{l}\right]=-\frac{1}{2}\,\bar{Q}_{l}\,,\qquad \left[F, Q^{k}\right]=\frac{1}{2}\,Q^{k}.
\eea
After introducing the new basis in this extension of \eqref{algebra-1},
\bea
    \tilde{H}= H - mF ,\label{H-tilde}
\eea
 one can pass to the centrally-extended superalgebra $\hat{su}(2|1)$ defined by the following non-vanishing (anti)commutators:
\bea
    &&\lbrace Q^{i}, \bar{Q}_{j}\rbrace = 2m I^i_j + 2\delta^i_j\left(H-mF\right) ,\qquad\left[I^i_j,  I^k_l\right]
    = \delta^k_j I^i_l - \delta^i_l I^k_j\,,\nn
    &&\left[I^i_j, \bar{Q}_{l}\right] = \frac{1}{2}\,\delta^i_j\bar{Q}_{l}-\delta^i_l\bar{Q}_{j}\, ,\qquad \left[I^i_j, Q^{k}\right]
    = \delta^k_j Q^{i} - \frac{1}{2}\,\delta^i_j Q^{k},\nn
    &&\left[F, \bar{Q}_{l}\right]=-\frac{1}{2}\,\bar{Q}_{l}\,,\qquad \left[F, Q^{k}\right]=\frac{1}{2}\,Q^{k}.\label{algebra}
\eea
Its bosonic sector contains the central charge generator $H$ (commuting with all other generators) and
the $U(2)_{\rm int}$ generators $I^i_j$ and $F$. Just this superalgebra was our point of departure in  \cite{DSQM}.

\subsection{Harmonic $SU(2|1)$ superspace as a coset superspace}
Harmonic $d=1$ superspace for the supergroup $\widehat{SU}(2|1)$ as a coset manifold of the latter can be defined using
the same tricks as in the construction of the harmonic superspace for the ${\cal N}=2, d=4$ superconformal group in \cite{HSS}.
The basic steps are passing to the new basis in \p{algebra}, in which all generators are labeled by their $U(1)$ charges, and
introducing an extra automorphism group $SU(2)_{\rm ext}$ which uniformly rotates all the doublet indices of the generators
in  \p{algebra}.

Using the notations
\bea
    &&Q^1 \equiv Q^+,\quad Q^2 \equiv Q^-,\quad\bar{Q}_1 \equiv \bar{Q}^-,\quad\bar{Q}_2 \equiv -\bar{Q}^+,\nn
    &&I^{++}\equiv I^1_2\,,\quad I^{--}\equiv I^2_1\,,\quad I^{0}\equiv I^1_1-I_2^2 =2I^1_1\,,\label{sub+}
\eea
we can rewrite the relations \eqref{algebra} as
\bea
    &&\lbrace Q^{-}, \bar{Q}^{+}\rbrace = m I^{0} - 2 H + 2 m F,\qquad\lbrace Q^{+}, \bar{Q}^{-}\rbrace = m I^{0} + 2 H - 2 m F,\nn
    &&\lbrace Q^{\pm}, \bar{Q}^{\pm}\rbrace = \mp\,2m I^{\pm\pm} ,\qquad \left[I^0,I^{\pm\pm}\right]=\pm\, 2 I^{\pm\pm},\quad\left[I^{++},I^{--}\right]=I^0,\nn
    &&\left[I^0, \bar{Q}^{\pm}\right] = \pm\,\bar{Q}^{\pm},\qquad\left[I^{++}, \bar{Q}^{-}\right]=\bar{Q}^{+},\qquad\left[I^{--}, \bar{Q}^{+}\right]=\bar{Q}^{-},\nn
    &&\left[I^0, Q^{\pm}\right]= \pm\,Q^{\pm},\qquad \left[I^{++}, Q^{-}\right]=Q^{+},\qquad \left[I^{--}, Q^{+}\right]=Q^{-},\nn
    &&\left[F, \bar{Q}^{\pm}\right]=-\frac{1}{2}\,\bar{Q}^{\pm}\,,\qquad \left[F, Q^{\pm}\right]=\frac{1}{2}\,Q^{\pm}.\label{algebra+}
\eea

We can also add, to the $su(2|1)$ superalgebra \eqref{algebra+}, the automorphism group $SU(2)_{\rm ext}$ with
the generators $\{T^0,T^{++},T^{--}\}$ which rotate the supercharges in the precisely same way as the internal $SU(2)_{\rm int}$ generators $\{I^0,I^{++},I^{--}\}$ do.
For consistency, the  $SU(2)_{\rm ext}$ generators should rotate, in the same way, the indices of
the $SU(2)_{\rm int}$ generators $I^i_j$, so these two $SU(2)$ groups form a semi-direct product
\bea
    \left[T,I\right] \propto I .
\eea

Then we introduce the following harmonic coset of the extended supergroup:
\be
\frac{\widehat{SU}(2|1)\rtimes SU(2)_{\rm ext}}{U(1)_{\rm int}\times U(1)_{\rm ext}} =
\frac{\{H, Q^i, \bar Q_k, F, I^i_j, T^{\pm\pm}, T^0\}}{\{F, T^0\}} \sim \left(t, \theta_i, \bar\theta^k, v^j_i, u^\pm_j\right),
\ee
where the harmonic variables $v_i^j$ parametrize the group $SU(2)_{\rm int}$, while $u^\pm_i, u^{+i}u^-_i =1\,, $ are the standard harmonics
on the coset $SU(2)_{\rm ext}/U(1)_{\rm ext}\sim S^2$ \cite{HSS0}. As the next step, one passes to the ``minimal'' complex harmonic coset
\be
\frac{\{H, Q^\pm, \bar Q^\pm, F, I^{\pm\pm}, I^0,  T^{\pm\pm}, T^0\}}{\{F, I^{++}, I^0, I^{--} - T^{--}, T^0\}} \sim
\left(t_{(A)}, \theta^\pm, \bar\theta^\pm, w^\pm_i\right) =: \zeta_H\,.\label{HB}
\ee
It is a deformation of the standard ``flat'' ${\cal N}=4$, $d=1$ harmonic superspace \cite{IvLe}. The new harmonics $w^\pm_i$
still satisfy the standard condition $w^{+ i}w^-_i = 1$ and are properly constructed from the bi-harmonic set $(v^i_j, u^\pm_k)$ \cite{HSS}.

We skip most details of the whole construction which basically coincide with those given in Sect. 9.1 of the book \cite{HSS}.
The set of coordinates defined in \p{HB} will be referred to as the {\it analytic basis} of the $SU(2|1)$ harmonic superspace.
The odd $SU(2|1)$ transformations in this basis obtained as left shifts of the relevant supercoset are written as
\bea
    &&\delta \theta^{+}= \epsilon^{+}+ m\,\bar{\theta}^{+}\theta^{+}\epsilon^{-}, \qquad
    \delta \bar{\theta}^{+}= \bar{\epsilon}^{+} - m\,\bar{\theta}^{+}\theta^{+}\bar{\epsilon}^{-},\nn
    &&\delta \theta^{-}= \epsilon^{-}+ 2m\,\bar{\epsilon}^{-}\theta^{-} \theta^{+}, \qquad
    \delta \bar{\theta}^{-}= \bar{\epsilon}^{-} + 2m\,\epsilon^{-}\bar{\theta}^{-}\bar{\theta}^{+},\nn
    &&\delta w^+_{i}=-\,m\left(\bar{\theta}^{+}\epsilon^{+}+\theta^{+} \bar{\epsilon}^{+} \right)w^-_{i},\qquad \delta w^-_{i}=0\,,\qquad
    \delta t_{(A)} = 2 i \left( \epsilon^{-}\bar{\theta}^{+}+\theta^{+} \bar{\epsilon}^{-}\right),\label{HBtr}
\eea
where
\bea
   \epsilon^{\pm} := \epsilon^{i}w^{\pm}_{i}\,, \qquad \bar{\epsilon}^{\pm} = \bar{\epsilon}^{k} w^{\pm}_k\,, \quad w^+_iw^-_k - w^+_k w^-_i = \varepsilon_{ik}\,.
\eea
Notice the asymmetry in the  transformations of the harmonic variables $w^+_i$ and $w^-_k$. This is a general feature of the
harmonic extensions of curved superspaces \cite{sg2,HSS}, and it was already encountered in the $d=1$ harmonic superspace
formalism, when considering realization of the ${\cal N}=4, d=1$ superconformal group $D(2,1;\alpha)$ on the harmonic coordinates  \cite{IvLe}.

It follows from the transformations \p{HBtr} that the $SU(2|1)$ harmonic superspace contains the analytic harmonic
subspace parametrized by the reduced coordinate set
\bea
    \zeta_{(A)} :=\left( t_{(A)}, \bar{\theta}^{+}, \theta^{+}, w^{\pm}_i\right),\lb{AnalSubs}
\eea
which is closed under the action of $SU(2|1)$. It can be identified with the supercoset
\footnote{By passing to the new basis in the semi-direct product of two $SU(2)$ groups involved in \p{HB}, \p{AnalCos}, $(I, T)
\rightarrow (I, {\cal J} = I - T)$, where the generators ${\cal J}^i_j$ commute with all other ones including $I^i_j$,
the supercoset \p{AnalCos} is split into the product
\bea
\frac{\{H, Q^\pm, \bar Q^\pm, F, I^{\pm\pm}, I^0,  {\cal J}^{\pm\pm}, {\cal J}^0\}}{\{Q^+, \bar Q^+, F, I^{++}, I^0, {\cal J}^{--}, {\cal J}^0\}}
= \frac{\{H, Q^\pm, \bar Q^\pm, F, I^{\pm\pm}, I^0 \}}{\{Q^+, \bar Q^+, F, I^{++}, I^0\}} \otimes \frac{\{{\cal J}^{\pm\pm}, {\cal J}^0\}}
{\{{\cal J}^{--}, {\cal J}^0\}}\,.  \nonumber
\eea
The first factor stands for the pure $SU(2|1)$ harmonic supercoset \cite{DSQM}  in which the internal part
is the complex coset of $SU(2)_{\rm int}$ over its  {\it parabolic} subgroup and which can be parametrized
by $(t_{(A)}, \theta^+, \bar\theta^+, \lambda^{++})$ where $\lambda^{++}$ is a complex ($\mathbb{CP}^1$) coordinate of the internal space.
The second factor fully decouples since the generators involved in its definition commute with the $SU(2|1)$ ones. While such
a ``minimal'' definition of the harmonic analytic $SU(2|1)$ superspace is in the spirit of the approach of refs. \cite{Howe}, here we prefer
to represent the internal sector by the harmonic variables $w^{\pm}_i$, like in \cite{HSS}.}
\bea
\frac{\{H, Q^\pm, \bar Q^\pm, F, I^{\pm\pm}, I^0,  T^{\pm\pm}, T^0\}}{\{Q^+, \bar Q^+, F, I^{++}, I^0, I^{--} - T^{--}, T^0\}}  \sim  \zeta_{(A)}\,.
\lb{AnalCos}
\eea

One can define the analytic subspace integration measure
\bea
d\zeta_{(A)}^{--} := dw\, dt_{(A)} \,d\bar{\theta}^{+}d\theta^{+},
\eea
which is invariant under the supersymmetry transformations \eqref{HBtr}.
The corresponding full integration measure $d\zeta_H$ in the analytic basis can be written as
\bea
d\zeta_H := dw\, dt_{(A)} \,d\bar{\theta}^{-}d\theta^- d\bar{\theta}^{+}d\theta^{+} \left(1 + m\,\bar{\theta}^{+}\theta^{-}
-m\, \bar{\theta}^{-}\theta^{+}\right),\lb{Fullmeas}
\eea
and it transforms as
\bea
\delta (d\zeta_H)=d\zeta_H\left[-\,m\left(\bar{\theta}^{-}\epsilon^{+}+\theta^{-}\bar{\epsilon}^{+}\right)
\left(1 - m\,\bar{\theta}^{+}\theta^{-}+m\, \bar{\theta}^{-}\theta^{+}\right) -m\left(\bar{\theta}^{+}\epsilon^{-}+\theta^{+}\bar{\epsilon}^{-} \right)\right].
\label{ZetaHtr}
\eea
One can check that there is no way to achieve the $SU(2|1)$ invariance of this measure: no a scalar factor can be picked up to compensate
the non-zero variation \p{ZetaHtr}. The relation of the analytic basis \p{HB} to the central basis containing the standard $SU(2|1)$ superspace \cite{DSQM}
is described in the Appendix B. In particular, it is shown that the measure \p{Fullmeas} in the central basis is reduced  to the product of the standard invariant
integration measure of the $SU(2|1)$ superspace and the non-invariant harmonic measure.

\subsection{Covariant derivatives}
We use the standard notation for the partial harmonic derivatives
\bea
&&    \partial^{\pm\pm} := w^\pm_i \frac{\partial}{\partial w^\mp_i} \,,
    \qquad  \partial^0 := w^+_i \frac{\partial}{\partial w^+_i} - w^-_i \frac{\partial}{\partial w^-_i}\,,\lb{HarmDer} \\
&&    [\partial^{++}, \partial^{--}] = \partial^0\,, \quad [\partial^0, \partial^{\pm\pm}] = \pm\,2 \partial^{\pm\pm}\,.\lb{HrmComm}
\eea
The deformed covariant spinor and harmonic derivatives can be derived by the routine coset techniques applied to the supercoset \p{HB}.
Once again, we skip details  and present the answer
\bea
    {\cal D}^{-} &=& -\frac{\partial}{\partial \theta^+} - 2i \,\bar\theta^-\partial_{(A)} +2 m\,\bar\theta^- \tilde{F}
    -m\,\bar{\theta}^{-}\left(\theta^+\frac{\partial}{\partial \theta^+}
+ \bar\theta^+\frac{\partial}{\partial \bar\theta^+} \right)-m\,\bar{\theta}^{-}\partial^0
    +m\,\bar{\theta}^{+}\partial^{--} ,\nn
    \bar{{\cal D}}^{-} &=& \frac{\partial}{\partial \bar\theta^+} - 2i\, \theta^- \partial_{(A)} +2 m\,\theta^- \tilde{F}+m\,
    \theta^{-}\left(\theta^+\frac{\partial}{\partial \theta^+}
+ \bar\theta^+\frac{\partial}{\partial \bar\theta^+} \right)+m\,\theta^{-}\partial^0
    -m\,\theta^{+}\partial^{--} ,\nn
    {\cal D}^{+} &=& \frac{\partial}{\partial \theta^-}+m\,\bar{\theta}^{-}\left(1+m\,\theta^{-} \bar{\theta}^{+} \right) \partial^{++}
    + 2im\,\bar{\theta}^{-} \theta^{+}\bar\theta^+\partial_{(A)} - 2 m^2\,\bar{\theta}^{-}\theta^{+}\bar\theta^{+}\tilde{F} \nn
    &&+\,m\,\bar{\theta}^{-} \theta^{+}\frac{\partial}{\partial \theta^{-}} +m\,\bar{\theta}^{-} \bar\theta^{+} \frac{\partial}{\partial \bar\theta^{-}}\,,\nn
    \bar{{\cal D}}^{+} &=& -\frac{\partial}{\partial \bar\theta^-}-m\,\theta^{-}\left(1-m\,\theta^{+} \bar{\theta}^{-} \right) \partial^{++}
    - 2im\,\theta^{-} \theta^{+}\bar\theta^+\partial_{(A)} + 2 m^2\,\theta^{-}\theta^{+}\bar\theta^{+}\tilde{F} \nn
    &&- \,m\,\theta^{-} \theta^{+}\frac{\partial}{\partial \theta^{-}} -m\,\theta^{-} \bar\theta^{+} \frac{\partial}{\partial \bar\theta^{-}}\,,\lb{SinorDe}
\eea
\bea
    {\cal D}^{++} &=& \left(1 + m\,\bar{\theta}^{+}\theta^{-}-m\, \bar{\theta}^{-}\theta^{+}\right)^{-1} \partial^{++}
    + 2i \,\theta^+\bar\theta^+\partial_{(A)} - 2 m\,\theta^+\bar\theta^{+}\tilde{F}
    + \theta^{+}\frac{\partial}{\partial \theta^{-}} + \bar\theta^{+}\frac{\partial}{\partial \bar\theta^{-}}\,,\nn
    {\cal D}^{--} &=& \left(1 + m\,\bar{\theta}^{+}\theta^{-}-m\, \bar{\theta}^{-}\theta^{+}\right)\partial^{--}
    + 2i \,\theta^{-}\bar\theta^{-}\partial_{(A)} - 2 m\,\theta^{-} \bar\theta^{-}\tilde{F}
+ \theta^{-}\frac{\partial}{\partial \theta^+} +
\bar\theta^{-}\frac{\partial}{\partial \bar\theta^{+}}\,,\nn
    {\cal D}^0 &=& \partial^0 + \left(\theta^+\frac{\partial}{\partial \theta^+}
+ \bar\theta^+\frac{\partial}{\partial \bar\theta^{+}} \right)
- \left(\theta^{-}\frac{\partial}{\partial \theta^{-}} +
\bar\theta^{-}\frac{\partial}{\partial \bar\theta^{-}} \right),\nn
    {\cal D}_{(A)} &=& \partial_{(A)}\,, \qquad \partial_{(A)}=\frac{\partial}{\partial t_{(A)}}\,.\label{cov}
\eea
Here, $\tilde{F}$ is a matrix part of the $U(1)_{\rm int}$ generator $F$. One can check that
these derivatives are indeed covariant under the transformations \p{HBtr}. The (anti)commutation relations among them
mimic those of the superalgebra \eqref{algebra+}:
\bea
    &&\{\bar{{\cal D}}^{+}, {\cal D}^{-}\}=m{\cal D}^0 - 2m\tilde{F} + 2i{\cal D}_{(A)}\,, \qquad\{\bar{{\cal D}}^{-}, {\cal D}^{+}\}
    =m{\cal D}^0 + 2m\tilde{F} - 2i{\cal D}_{(A)}\,,\nn
    &&\{{\cal D}^{\pm}, \bar{{\cal D}}^{\pm}\}=\mp \,2m {\cal D}^{\pm\pm}, \qquad\left[{\cal D}^{++}, {\cal D}^{--}\right] = {\cal D}^0,
    \qquad \left[{\cal D}^0, {\cal D}^{\pm\pm}\right] = \pm \, 2 {\cal D}^{\pm\pm},\nn
    &&\left[{\cal D}^{++}, {\cal D}^{-}\right]={\cal D}^{+}, \qquad \left[{\cal D}^{--}, {\cal D}^{+}\right]
    ={\cal D}^{-},\qquad \left[{\cal D}^{0}, {\cal D}^{\pm}\right]=\pm\,{\cal D}^{\pm},\nn
    &&\left[{\cal D}^{++}, \bar{{\cal D}}^{-}\right]=\bar{{\cal D}}^{+}, \qquad \left[{\cal D}^{--},\bar{{\cal D}}^{+}\right]=\bar{{\cal D}}^{-},
    \qquad \left[{\cal D}^{0}, \bar{{\cal D}}^{\pm}\right]= \pm\,\bar{{\cal D}}^{\pm},\lb{covcomm}\\
    &&\tilde{F}\, {\cal D}^{\pm}=-\frac{1}{2}\,{\cal D}^{\pm}, \qquad   \tilde{F}\, \bar{{\cal D}}^{\pm} = \frac{1}{2}\,\bar{{\cal D}}^{\pm}. \lb{Fcomm}
\eea

It will be  convenient to represent the derivatives ${\cal D}^{+}\,$, $\bar{{\cal D}}^{+}$  as
\bea
    {\cal D}^{+}=\frac{\partial}{\partial \theta^-}+m\,\bar{\theta}^{-}{\cal D}^{++}, \qquad
    \bar{{\cal D}}^{+}= -\frac{\partial}{\partial \bar\theta^-}-m\,\theta^{-}{\cal D}^{++}.
\eea
These spinor derivatives, together with ${\cal D}^{++}$ and ${\cal D}^0$, form the so-called CR (``Cauchy-Riemann'') structure \cite{HSS}
\bea
&& \{{\cal D}^+,  \bar{{\cal D}}^{+}\} = -\,2m\,{\cal D}^{++}, \; \{{\cal D}^+,  {\cal D}^{+}\} = \{\bar{{\cal D}}^+,  \bar{{\cal D}}^{+}\} = 0\,, \;
\left[{\cal D}^{++},  {\cal D}^+ \right] = \left[{\cal D}^{++},  \bar{{\cal D}}^+ \right] = 0\,, \nn
&& \left[{\cal D}^0, {\cal D}^+\right] = {\cal D}^+\,, \quad \left[{\cal D}^0, \bar{\cal D}^+\right] = \bar{\cal D}^+\,, \quad
\left[{\cal D}^0, {\cal D}^{++}\right] = 2\,{\cal D}^{++}\,. \lb{CRstruct}
\eea
All other (anti)commutators can be derived from these ones by applying the harmonic derivative ${\cal D}^{--}$ which, together with
${\cal D}^{++}$ and ${\cal D}^0$, form an $SU(2)$ algebra. The non-standard feature of the considered case
is that the analyticity-preserving covariant harmonic derivative ${\cal D}^{++}$ in the CR structure \p{CRstruct} does not decouple
from the spinorial derivatives ${\cal D}^+$, $\bar{{\cal D}}^{+}$. This will entail the essential differences of the $SU(2|1)$ harmonic
formalism and the relevant $SU(2|1)$ mechanics models from their flat ${\cal N}=4, d=1$ harmonic superspace prototypes. Another
peculiarity of the  $SU(2|1)$ harmonic formalism is the presence of the additional matrix $U(1)$ charge $\tilde{F}$
with the non-trivial action \p{Fcomm} on the spinor covariant derivatives. One should be careful about taking it correctly into account, while
checking various (anti)commutators, in particular those in \p{CRstruct} \footnote{The correct result is still obtained, if
the l.h.s. in \p{CRstruct} is formally replaced by the commutators. However, one should remember that the definite  $\tilde{F}$ charge is
ascribed to a covariant spinor derivative as a whole, not to its separate constituents.}.

\subsection{Harmonic $SU(2|1)$ superfields}
The passive odd transformation of the harmonic superfields in the analytic basis $\Phi\left(\zeta_H\right)$ can be written as
\bea
    \delta \Phi = -\,m\left[2 \left(\bar{\theta}^{+}\epsilon^{-} - \theta^{+}\bar{\epsilon}^{-}\right) \tilde{F} + \left(\bar{\theta}^{+}\epsilon^{-}
    + \theta^{+}\bar{\epsilon}^{-}\right) {\cal D}^{0}
    +\left(\bar{\theta}^{-}\epsilon^{-}+\theta^{-} \bar{\epsilon}^{-} \right){\cal D}^{++}\right]\Phi. \label{SFtr}
\eea
The superfields $\Phi$ are assumed to have definite $U(1)$ charges, $\tilde{F}\Phi = \kappa \Phi\,, \;{\cal D}^0\Phi = q\,\Phi$\,.
The presence of the derivative ${\cal D}^{++}$ in \p{SFtr} is necessary for the correct $SU(2|1)$ closure of these variations
and for ensuring that various covariant derivatives of $\Phi$, e.g., ${\cal D}^{++}\Phi, \;{\cal D}^+\Phi$ and ${\cal D}^{--}\Phi\,$,
transform according to the same generic rule \p{SFtr} as $\Phi$ itself (see Appendix A).

Given some set of such superfields $\{\Phi_1, \Phi_2, \ldots \Phi_N\}$, we can write the general $\sigma$-model-type action as
\bea
    S =\int dt\,{\cal L}= \int d\zeta_H\, K\left(\Phi_1, \Phi_2, \ldots \Phi_N, w^{\pm}_i\right),\qquad {\cal D}^{0}
    K\left(\Phi_1, \Phi_2, \ldots \Phi_N, w^{\pm}_i\right)=0\,.\label{gen-A}
\eea
Here $K$ is a real function of superfields and the harmonic coordinates $w^{\pm}_{i}$, arbitrary for the moment.
Varying \eqref{gen-A} with respect to the supersymmetry transformations \eqref{HBtr}, \eqref{SFtr}, we obtain
\bea
    \delta S &=& \int d\zeta_H \,{\cal D}^{++}\Big\lbrace -m \left(\bar{\theta}^{-}\epsilon^{-}+ \theta^{-}\bar{\epsilon}^{-}\right)
    K\left(\Phi_1\ldots \Phi_N,w^{\pm}_i\right)\Big\rbrace\nn
    &&+\,\int d\zeta_H \, \Big\lbrace m\left(\bar{\theta}^{-}\epsilon^{-}+ \theta^{-}\bar{\epsilon}^{-}\right)
    \left(1 + m\,\bar{\theta}^{+}\theta^{-}-m\,\bar{\theta}^{-}\theta^{+}\right)^{-1}\hat{\partial}^{++}
    +m\left(\bar{\theta}^{+}\epsilon^{-}+\theta^{+}\bar{\epsilon}^{-}\right) \hat{\partial}^{0}\nn
    && -\,2m\left(\bar{\theta}^{+}\epsilon^{-} - \theta^{+}\bar{\epsilon}^{-}\right) \tilde{F}
    -m\left(\bar{\theta}^{+}\epsilon^{+}+\theta^{+} \bar{\epsilon}^{+} \right)\hat{\partial}^{--}\Big\rbrace\,K\left(\Phi_1\ldots \Phi_N,w^{\pm}_i\right), \lb{varK}
\eea
where $\hat{\partial}^\pm, \hat{\partial}^0$ act only on the explicit harmonics in the function $K$. Requiring the variation \p{varK} to vanish
up to a total derivative under the integral yields the following restrictions on $K$:
\bea
    &&\hat{\partial}^{\pm\pm}K\left(\Phi_1, \Phi_2, \ldots \Phi_N, w^{\pm}_i\right)=0\quad\Rightarrow\quad K=K\left(\Phi_1, \Phi_2, \ldots \Phi_N\right),\nn
    &&\tilde{F}K\left(\Phi_1, \Phi_2, \ldots \Phi_N\right)={\cal D}^{0}K\left(\Phi_1, \Phi_2, \ldots \Phi_N\right)=0\,.\label{SF}
\eea
Thus the general action is given by
\bea
    S =\int dt\,{\cal L}= \int d\zeta_H\, K\left(\Phi_1, \Phi_2, \ldots \Phi_N \right)\label{kin}
\eea
and its variation $\delta S$ is
\bea
    \delta S = \int d\zeta_H \,{\cal D}^{++}\Big\lbrace -m \left(\bar{\theta}^{-}\epsilon^{-}+ \theta^{-}\bar{\epsilon}^{-}\right)
    K\left(\Phi_1\ldots \Phi_N\right)\Big\rbrace=0\,.
\eea

In what follows we will be interested in the analytic superfields, i.e. those subjected to the Grassmann Cauchy-Riemann constraints
\bea
{\cal D}^{+}\phi =\bar{\cal D}^{+}\phi = 0\,. \lb{CRsuperf}
\eea
In virtue of the CR structure relations \p{CRstruct}, these analyticity constraints imply, as their integrability
condition, the harmonic Cauchy-Riemann condition
\bea
{\cal D}^{++}\phi =0\,.\label{A-cond}
\eea
This is in a crucial difference from the flat ${\cal N}=4, d=1$ harmonic analytic superfields \cite{IvLe} for which \p{CRsuperf}  do not
necessarily imply the appropriate version of \p{A-cond}. With taking into account \p{A-cond}, the derivatives ${\cal D}^{+}$ and $\bar{\cal D}^{+}$
become ``short'',
\be
\left({\cal D}^{+}\,, \;\bar{\cal D}^{+}\right) \; \Rightarrow \; \left(\frac{\partial}{\partial \theta^-}\,, \;
-\frac{\partial}{\partial \bar\theta^-}\right), \lb{Rightarrow}
\ee
and so \p{CRsuperf} give
\be
\phi = \phi(\zeta_{(A)})\,.
\ee
The necessity of the harmonic constraint \p{A-cond} for the preservation of Grassmann $SU(2|1)$ analyticity also directly follows from the
general superfield transformation law \p{SFtr}. Only under this constraint the variation of the analytic superfield does not contain
the coordinates $\theta^-, \bar\theta^-$.

Finally, we note that the analytic superspace \p{AnalSubs} can be extended to the two  mutually conjugated three-theta analytic superspaces
\bea
\zeta_{(A)}^{(3)} := \left(\theta^-, \zeta_{(A)}\right) = \left( t_{(A)}, \theta^-, \bar{\theta}^{+}, \theta^{+}, w^{\pm}_i\right)\,, \qquad
\bar\zeta_{(A)}^{(3)} = \left(\bar\theta^-, \zeta_{(A)}\right), \lb{AnalSubs3}
\eea
which are also closed under the coordinate transformations \p{HBtr}. The relevant superfields are singled out by the covariant chirality-like
conditions \footnote{These conditions are solved in terms of unconstrained superfields living on $\zeta_{(A)}^{(3)}$ or $\bar\zeta_{(A)}^{(3)}$ by
passing to the new frames where $\bar{\cal D}^+$ or ${\cal D}^+$ become short. This passing is accomplished by means of   the appropriate invertible
intertwining operators, e.g., ${\cal D}^+( 1 - m\,\theta^-\bar\theta^-{\cal D}^{++}) =
( 1 - m\,\theta^-\bar\theta^-{\cal D}^{++})\frac{\partial}{\partial \theta^-}$\,.}
\be
\bar{\cal D}^+\tilde{\phi}_{(1)} = 0 \quad {\rm or} \quad {\cal D}^+\tilde{\phi}_{(2)} = 0\,,
\ee
which do not require the harmonic constraints \p{A-cond}. The existence of analogous extended analytic superspaces
in the flat ${\cal N}=4, d=1$ case was noticed in \cite{IvNied}. Possible implications of these additional Grassmann
analyticities in the $SU(2|1)$ SQM models will be addressed elsewhere.

\section{The multiplet ${\bf (4,4,0)}$}
In the flat ${\cal N}=4, d=1$ supersymmetry the multiplet with the field contents ${\bf (4,4,0)}$ is described by an analytic harmonic superfield and
it is the basic ${\cal N}=4, d=1$ multiplet: all other irreducible
multiplets and the related SQM models can be obtained from this multiplet and the SQM models associated with it through different versions
of the Hamiltonian reduction \cite{root} or, equivalently, by gauging some isometries of the ${\bf (4,4,0)}$ Lagrangians \cite{DeldIv1,DeldIv2,DeldIv3}.

The $SU(2|1)$ version of the multiplet ${\bf (4,4,0)}$ is described by the superfield $q^{+a}$ satisfying the constraints
\be
    \bar{\cal D}^{+} q^{+a} ={\cal D}^{+} q^{+a} ={\cal D}^{++} q^{+a} = 0\,,\qquad \tilde{F}q^{+a} = 0\,.\label{q-constr}
\ee
These constraints look just as those in the flat ${\cal N}=4, d=1$ superspace (except for the last one). Their solution reads
\be
q^{+a}\left(\zeta_{(A)}\right) = x^{ia}w^{+}_i + \theta^{+} \psi^a +
\bar\theta^{+} \bar\psi^a -
2i \theta^{+}\bar{\theta}^{+} \dot{x}^{ia} w^{-}_i\,, \label{q+}
\ee
where
\be
\left(x^{ia}\right)^\dagger = \epsilon_{ab}\epsilon_{ik} x^{kb}\,,
\qquad \left(\psi^a\right)^\dagger = \bar{\psi}_a\,.
\ee
The index $a=1,2$ is the doublet index of the  ``Pauli-G\"ursey'' group $SU(2)_{PG}$ which commutes with $SU(2|1)$.
The fermionic fields $\psi^a, \bar{\psi}^a$ can be combined into a doublet of the external group $SU^{\prime}(2)$ as $(\psi^a, \bar{\psi}^a) :=
\psi^{i^{\prime}a}$.

The analytic superfield $q^{+a}$ has no dependence on $m$ in its $\theta$-expansion, however the non-analytic counterpart of
$q^{+a}$, i.e.  $q^{-a} :={\cal D}^{--}q^{+a}\,,$
displays such a dependence:
\bea
    q^{-a} &=& \left[1 + m\,\bar{\theta}^{+}\theta^{-}- m\,\bar{\theta}^{-}\theta^{+}\right]x^{ia}w^{-}_{i} + \theta^{-} \psi^{a} + \bar\theta^{-} \bar\psi^{a}
     +2i\left(\bar{\theta}^{+}\theta^{-}+\bar{\theta}^{-}\theta^{+}\right)\dot{x}^{ia} w^{-}_{i}\nn
     &&+\,2i\theta^{-}\bar{\theta}^{-}\left[\dot{x}^{ia}w^{+}_{i} + \theta^{+} \dot{\psi}^a +
\bar\theta^{+} \dot{\bar\psi}^a - 2i \theta^{+}\bar{\theta}^{+} \ddot{x}^{ia} w^{-}_{i}\right].
\eea

The odd $SU(2|1)$ transformation of $q^{+a}$ is a particular case of the general transformation law \eqref{SFtr},
\bea
    \delta q^{+a} = -\,m\left(\bar{\theta}^{+}\epsilon^{-} + \theta^{+}\bar{\epsilon}^{-}\right)q^{+a}.
\eea
It implies the following transformations for the component fields
\bea
    \delta x^{ia} =-\,\epsilon^i\psi^a -\bar{\epsilon}^i \bar{\psi}^a ,\qquad
    \delta \bar{\psi}_a =2i\epsilon_k \dot{x}^{k}_{a}- m\,\epsilon_k x^{k}_{a}\,,\qquad \delta \psi^a =2i\bar{\epsilon}^k \dot{x}^{a}_k+
    m\,\bar{\epsilon}^k x^{a}_k\,.\label{Hypertr}
\eea

Note that the matrix $U(1)$ generator $\tilde{F}$ can be ``activated'' on $q^{+ a}$ by identifying it with  some $U(1)\subset SU(2)_{\rm PG}$,
e.g., as
\be
\tilde{F} q^{+ a} = \kappa (\tau_3)^{\; a}_{b} q^{+ b}\,,
\ee
where $\kappa$ is a new charge. The formulas given above are generalized to the $\kappa \neq 0$ case as
\be
    q^{+a} = x^{ia}w^{+}_i + \theta^+ \psi^a +\bar\theta^{+} \bar\psi^a - 2i\,\theta^{+}\bar{\theta}^{+} \nabla_{(t)} x^{ia}w^{-}_i,
\ee
where
\be
\nabla_{(t)} x^{ia} := \dot{x}^{ia} + i\kappa m \left(\tau_3\right)^a_{\;b}{x}^{ib}\,.
\ee
According to the transformation law \eqref{SFtr},  $q^{+a}$ transforms as
\bea
    \delta q^{+a} = - m\left(\bar{\theta}^{+}\epsilon^{-} + \theta^{+}\bar{\epsilon}^{-}\right)q^{+a}
    - 2\kappa m \left(\bar{\theta}^{+}\epsilon^{-} - \theta^{+}\bar{\epsilon}^{-}\right)\left(\tau_3\right)^a_b q^{+b}.
\eea
The component fields transformations \eqref{Hypertr} are modified as
\bea
    &&\delta x^{ia} =-\,\epsilon^i\psi^a  -\bar{\epsilon}^i \bar{\psi}^a , \qquad \delta \psi^a =2 i\,\bar{\epsilon}^k \nabla_{(t)}x^{a}_{k}+ m\,\bar{\epsilon}^k x^{a}_k\,,\nn
    &&\delta \bar{\psi}_a =2 i\,\epsilon_k \nabla_{(t)}x^{k}_{a} - m\,\epsilon_k x^{k}_{a}\,.
\eea
\subsection{The general $\sigma$-model action}
For simplicity, we will basically deal with the $\kappa = 0$ case, leaving a comment
on the $\kappa \neq 0$ case for the end of this subsection.

In accord with the general structure of the harmonic superfield actions \eqref{kin}, the $\sigma$-model type action for the
multiplet $({\bf 4, 4, 0})\,$ can be written as
\bea
    S\left(q^{\pm a}\right)=\int d\zeta_H\,K\left(q^{+}, q^-\right),\label{kin-q}
\eea
where $K$ satisfies the conditions
\bea
    \tilde{F}K\left(q^{\pm a}\right)={\cal D}^{0}K\left(q^{\pm a}\right)=0\,.
\eea

Note that the essential difference of \p{kin-q} from its flat ${\cal N}=4$ counterpart \cite{IvLe}
is that the Lagrangian function $K$ cannot involve explicit harmonics (see \p{SF}).
Since the function $K$ is neutral, it can be written as a function of two neutral superfield arguments, $K=K\left(q^{2},X^{(ab)}\right)$,
with
\bea
&& q^2=2\,q^{+a}q^{-}_{a},\qquad X^{(ab)}=q^{+(a}q^{-b)}=\frac{1}{2}\left(q^{+a}q^{-b}+q^{+b}q^{-a}\right), \lb{SupArg} \\
&& {\cal D}^{++} q^2=\left({\cal D}^{++}\right)^2 X^{(ab)}=0\,,\qquad {\cal D}^{0}q^2={\cal D}^{0}X^{(ab)}=0\,.
\eea
The function $K\left(q^{2},X^{(ab)}\right)$ can be represented as a power series in $X^{(ab)}$:
\bea
    K\left(q^{2},X^{(ab)}\right)=K_0\left(q^{2}\right)+\sum_{n=1}^{\infty}
    C_{a_1 a_2\ldots a_n b_1 b_2\ldots b_n}\left(q^{2}\right)X^{(a_1 b_1)}X^{(a_2 b_2)}\ldots X^{(a_n b_n)}.\label{sum}
\eea
It can be shown that every term of this expansion, except for the zeroth order one $K_0\left(q^{2}\right)$,
is a total ${\cal D}^{++}$ derivative plus a function of $q^2$
which can be absorbed into $K_0\left(q^{2}\right)$. We explicitly show  this for $n=1,2$:
\bea
    &&X^{(a_1 b_1)}=\frac{1}{2}\,{\cal D}^{++}\left[q^{-a_1}q^{-b_1}\right],\nn
    &&X^{(a_1 b_1)}X^{(a_2 b_2)}=\frac{1}{3}\,{\cal D}^{++}\left[q^{-a_1}q^{-b_1}X^{(a_2 b_2)}\right]-\frac{1}{96}\,\epsilon^{a_1 a_2}\epsilon^{b_1 b_2}\left(q^2\right)^2\,.
\eea

So we have $K\left(q^{2},X^{(ab)}\right)= - L\left(q^{2}\right)+{\cal D}^{++}L^{--}\left(q^{2},X^{(ab)}\right)$
and the general action \eqref{kin-q} can be rewritten as
\bea
    S(q^{2})=\int dt\,{\cal L} =  -\int d\zeta_H \, L\left(q^{2}\right),\qquad
    \tilde{F}L\left(q^{2}\right)={\cal D}^{0}L\left(q^{2}\right)=0\,,\label{Sq}
\eea
where the sign minus was chosen for further convenience. The corresponding component Lagrangian reads
\footnote{We use the following convention for Grassmann variables: $\left(\chi\right)^2 = \chi_i\chi^i$, $\left(\bar{\chi}\right)^2 = \bar{\chi}^i\bar{\chi}_i$\,.}
\bea
    {\cal L}&=& G\left[\dot{x}^{ia}\dot{x}_{ia} + \frac{i}{2}\left(\bar{\psi}_a\dot{\psi}^a-\dot{\bar{\psi}}_a\psi^a\right)+\frac{m}{2}\,\psi^a\bar{\psi}_a\right]
    -\frac{i}{2}\,\dot{x}^{ia}\partial_{ic}G\left(\psi_a\bar{\psi}^c+\psi^c\bar{\psi}_a\right)\nn
    &&-\,\frac{\Delta_x G}{16}\left(\bar\psi\,\right)^2\left(\psi\right)^2 +\frac{m}{4}\,x^{ic}\partial_{ic}G\,\psi^a\bar{\psi}_a
    - \frac{m^2}{4}\,x^2\,G\,,\lb{LagrKincomp}
\eea
where
\bea
    &&\partial_{ia}=\partial/\partial x^{ia},\qquad \Delta_x = \epsilon^{ik}\epsilon^{ab}\partial_{ia}\partial_{kb}\,,\qquad x^2=x^{ia}x_{ia}\,,\nn
    &&G\left(x^2\right) = \Delta_x L\left(x^2\right)= 4x^2\, L^{\prime\prime}\left(x^2\right) + 8 L^{\prime}\left(x^2\right), \;\partial_{ia}G =
    8x_{ia}\left(3L^{\prime\prime} + x^2L^{\prime\prime\prime}\right).
\eea
We observe that $SU(2|1)$ supersymmetry imposes rather severe restrictions on the structure of the ${\bf(4, 4, 0)}$ Lagrangian as compared with the standard
${\cal N}=4, d=1$ supersymmetry. Though the bosonic metric is conformally flat in both cases, in the $SU(2|1)$ case it turns out
to be $SU(2)_{\rm int}\times SU(2)_{\rm PG} \sim SO(4)$ symmetric, with the conformal factor being a function of $x^2$. The extra fermionic $SU^\prime(2)$ symmetry
is broken by the terms $\sim \psi^a\bar\psi_a = \psi^{1'a}\psi^{2'}_a$. Note that the whole action respects $SU(2)_{\rm PG}$ symmetry.

The simplest case corresponds to the free system with $G=1$:
\bea
&& S_{\rm free}(q^{\pm a}) = -\frac{1}{4} \int d\zeta_H\, q^{+a}q^{-}_{a}\,,\label{action-q+}\\
&&{\cal L}_{\rm free}= \dot{x}^{ia}\dot{x}_{ia} + \frac{i}{2}\left(\bar{\psi}_a\dot{\psi}^a-\dot{\bar{\psi}}_a\psi^a\right)+\frac{m}{2}\,\psi^a\bar{\psi}_a
- \frac{m^2}{4}\, x^{ia} x_{ia}\,.\label{Lqfree}
\eea

Let us make a brief comment concerning the case with non-zero external $\tilde{F}$ charge $\kappa \neq 0$ on the example of the free action.
The relevant modified Lagrangian can be obtained just via changing $\partial_t \rightarrow  \nabla_{(t)} = \partial_t + i\kappa m \tau_3$
in \p{Lqfree}, which yields
\bea
{\cal L}_{\rm free}^{(\kappa)} &=& \dot{x}^{ia}\dot{x}_{ia} + \frac{i}{2}\left(\bar{\psi}_a\dot{\psi}^a-\dot{\bar{\psi}}_a\psi^a\right)
- 2i\kappa m (\tau_3)^a_b \dot{x}^{ib} x_{ia} +\frac{m}{2}[ \delta^a_b + 2\kappa (\tau_3)^a_b]\,\psi^b\bar{\psi}_a \nn
&&+\, \left(\kappa^2 - \frac14 \right){m^2}\, x^{ia} x_{ia}\,.\label{Lqkappa}
\eea
Note the appearance of the induced WZ term $\sim \kappa$ and the additional contributions to the mass terms in this Lagrangian.
The $SU(2)_{\rm PG}$ symmetry gets broken at $\kappa \neq 0$.

\subsection{The absence of WZ type actions}
The most general WZ (or CS) action \cite{IvLe} is given by the integral over the analytic subspace
\bea
    S_{\rm WZ}(q^{+a}) =-\frac{i}{2} \int d\zeta^{--}_A \,L^{++}\left(q^{+a}, w^{\pm}_i\right).
\eea
Since the analytic superfield \eqref{q+} is not deformed by $m$,
this action coincides with the non-deformed WZ action for the multiplet ${\bf (4,4,0)}$ given in \cite{IvLe}.
The component Lagrangian reads
\bea
    {\cal L}_{\rm WZ} = {\cal A}_{ia}\dot{x}^{ia} - \frac{i}{2} \, {\cal B}_{(ab)} \psi^a \bar{\psi}^b,\label{WZq}
\eea
where
\bea
    {\cal A}_{ia}\left(x^{ia}\right):=\int dw\,w^{-}_i\,\frac{\partial L^{++}}{\partial x^{+a}}\,, \qquad
    {\cal B}_{(ab)}\left(x^{ia}\right):=\int dw\,\frac{\partial^2 L^{++}}{\partial x^{+a}\partial x^{+b}}\,.\label{gauge-0}
\eea
By construction, the external gauge field ${\cal A}_{ia}$ satisfies the self-duality condition
\be
    {\cal F}_{kb\,ia} := \partial_{kb} {\cal A}_{ia} - \partial_{ia} {\cal A}_{kb} =
\epsilon_{ki}\,\int du \,
\frac{\partial^2 L^{++}}{\partial x^{+a}\,\partial x^{+b}}
=: \epsilon_{ki} {\cal B}_{(ab)}\,, \label{selfd}
\ee
and the transversal gauge condition
\be
    \partial_{ia}{\cal A}^{ia} = 0\,.\label{gauge}
\ee
The Lagrangian \eqref{WZq} transforms under the $SU(2|1)$ transformations \eqref{Hypertr} as
\bea
    \delta{\cal L}_{\rm WZ}= -\,\partial_t\left[{\cal A}^{ia}\left(\epsilon^i\psi^a + \bar{\epsilon}^i \bar{\psi}^a\right)\right]
    +\frac{i}{2}\,m\,{\cal B}_{(ab)}\left(\epsilon^k\psi^a - \bar{\epsilon}^k\bar{\psi}^a \right)x^{b}_k\,,\lb{VarWZ1}
\eea
and this variation is not reduced to a total derivative because of the term $\sim m$. Thus,
the Lagrangian is not $SU(2|1)$ invariant for any choice of $L^{++}$, which just means the absence of the WZ action for the multiplet
${\bf (4, 4, 0)}$ in the case of $SU(2|1)$ supersymmetry. The same conclusion is valid for the $\kappa \neq 0$ case as well.

Some further issues regarding the superconformal properties of WZ term will be discussed in section \ref{section 5}.
\subsection{Hamiltonian formalism}
The classical Hamiltonian obtained as the Legendre transform  of the Lagrangian \p{LagrKincomp} reads
\bea
    H &=& \frac{1}{4G}\left[p^{ia}+\frac{i}{2}\left(\bar{\psi}_c\psi^a+\bar{\psi}^a\psi_c\right)\partial^{ic}G\right]
    \left[p_{ia}-\frac{i}{2}\left(\bar{\psi}_a\psi^b+\bar{\psi}^b\psi_a\right)\partial_{ib}G\right]\nn
    &&+\,\frac{\Delta_x G}{16}\left(\bar\psi\,\right)^2\left(\psi\right)^2 - \frac{m}{4}\left(2G+x^{ic}\partial_{ic}G\right)\psi^a\bar{\psi}_a
    + \frac{m^2}{4}\,x^{ia} x_{ia}\,G\, .
\eea
It is also straightforward to find the supercharges $\left(Q^i\right)^\dagger=\bar{Q}_i$, as well as the remaining bosonic generators,
\bea
    &&Q_i = \psi^a\left(p_{ia}+imx_{ia}G+\frac{i}{4}\,\psi_a\bar{\psi}^b\partial_{ib}G\right),\quad
    \bar{Q}_i = \bar{\psi}^a\left(p_{ia}-imx_{ia}G-\frac{i}{4}\,\bar{\psi}_a\psi^b\partial_{ib}G\right),\nn
    &&F = \frac{1}{2}\,G\,\psi^a\bar{\psi}_a\,,\qquad I_{ik} = i x^{a}_{(i}\,p_{k)a}\,.\lb{ClassQ}
\eea
The Poisson brackets for the bosonic variables and the Dirac brackets for fermions are defined as
\bea
    &&\lbrace p_{ia}, x^{kb}\rbrace = -\,\delta_i^k\delta_a^b \,,\qquad
    \lbrace \psi^a, \bar{\psi}_b\rbrace = -\,i\,G^{-1}\delta_b^a\,,\nn
    &&\lbrace p_{ia}, \psi^b\rbrace = \frac{1}{2}\,\psi^b G^{-1}\partial_{ia} G\,,\qquad
    \lbrace p_{ia}, \bar{\psi}_b\rbrace = \frac{1}{2}\,\bar{\psi}_b\, G^{-1}\partial_{ia} G\,.\label{brackets}
\eea
To simplify the brackets, it is useful to make the substitutions
\bea
    \psi^a = G^{-\frac{1}{2}}\xi^a,\qquad \bar{\psi}_b=G^{-\frac{1}{2}}\bar{\xi}_b\,,
\eea
whence
\bea
    \lbrace p_{ia}, x^{kb}\rbrace = -\,\delta_i^k\delta_a^b \,,\qquad
    \lbrace \xi^a, \bar{\xi}_b\rbrace = -\,i\,\delta_b^a\,,\qquad
    \lbrace p_{ia}, \xi^b\rbrace =  \lbrace p_{ia}, \bar{\xi}_b\rbrace =0\,.\label{brackets2}
\eea
These brackets can be quantized   in the standard way as
\bea
    p_{ia}=-\,i\partial_{ia}\,,\qquad \bar{\xi}_a = \partial/\partial \xi^a ,\qquad
    \left[p_{ia}, x^{kb}\right] = -\,i\,\delta_i^k\delta_a^b \,,\qquad \lbrace \xi^a, \bar{\xi}_b\rbrace = \delta_b^a\,.
\eea

The quantum supercharges can be constructed from the classical ones \p{ClassQ} according to the prescriptions of ref. \cite{How}
which were applied by us also in \cite{DSQM}. Their basic steps are Weyl-ordering and the subsequent similarity transformation defined
in terms of the target bosonic metric. In the present case this general procedure yields
\bea
    &&Q_{{\rm (cov)}i} = -iG^{-\frac{1}{2}} \xi^a\left[(\partial_{ia}- mx_{ia})G
    -\frac14 \left(\xi_a\bar{\xi}^b - 2\delta^b_a\right) G^{-1}\partial_{ib}G\right],\nn
    &&\bar{Q}_{{\rm (cov)}i} = -iG^{-\frac{1}{2}} \bar{\xi}^a\left[(\partial_{ia} + mx_{ia})G
    + \frac14\left(\bar{\xi}_a\xi^b + 2\delta^b_a\right)G^{-1}\partial_{ib}G\right].\lb{QuantumQ}
\eea
The quantum Hamiltonian is defined by the anticommutator of the quantum supercharges, and it reads
\bea
    H_{\rm (cov)} &=& - \frac{1}{4G}\left[\partial^{ia}
    +\frac{1}{2}\left(\xi^a\bar{\xi}_c - \bar{\xi}^a\xi_c\right)G^{-1}\partial^{ic}G\right]
    \left[\partial_{ia} + \frac{1}{2}\left(\bar{\xi}_a\xi^b -\xi_a\bar{\xi}^b\right)G^{-1}\partial_{ib}G\right]\nn
    &&+\,\frac{\Delta_x G}{16 G^2}\left[\left(\xi\right)^2\left(\bar\xi\,\right)^2 - 2\xi^a\bar\xi_a \right]
    - \frac{m}{8}\left(2+x^{ic}G^{-1}\partial_{ic}G\right)\left[\xi^a,\bar{\xi}_a\right]\nn
    &&+\, \frac{m^2}{4}\,x^{ia} x_{ia}\,G -\frac{m}{2}\,. \lb{QuantH}
\eea
The rest of the superalgebra generators is
\bea
F = -\frac{1}{2}\,\bar{\xi}_a\xi^a,\qquad I_{ik} = x^{a}_{(i}\,\partial_{k)a}\,.\label{qgen}
\eea
Taken together, these generators form the $\hat{su}(2|1)$ superalgebra \p{algebra}.

Note that the quantum covariantized supercharges can be conveniently represented as
\bea
    Q_{{\rm (cov)}i} = e^{-mW}Q_{{\rm (cov)}i}^{\left(m=0\right)}e^{mW},\qquad\bar{Q}_{{\rm (cov)}i}
    = e^{mW}\bar{Q}_{{\rm (cov)}i}^{\left(m=0\right)}e^{-mW},\label{W-rep}
\eea
where the function $W$ is given by the expression
\bea
    W(x^2)=2 x^2 L^\prime(x^2)+2L(x^2)\,,\quad \partial_{ia}W\left(x^2\right)
    =-\,x_{ia}G\left(x^2\right),\quad 2W^\prime\left(x^2\right) =-\,G\left(x^2\right). \lb{SimTr}
\eea
Thus the quantum $SU(2|1)$ supercharges for the multiplet ${\bf (4, 4, 0)}$ can be obtained  from their
flat ${\cal N}=4, d=1$ counterparts (with the special $SO(4)$ invariant target bosonic metric) through a similarity transformation. This
is a particular case of the general phenomenon summarized in \cite{Taming}. To avoid a possible confusion, we notice that the
transformation \p{W-rep} is not unitary, and for this reason the Hamiltonian \p{QuantH} is by no means equivalent to its $m=0$ counterpart.

\subsection{The free model: Spectrum and the $SU(2|1)$ Casimirs}
Let us consider the simplest case $G=1$ corresponding to the free Lagrangian \eqref{Lqfree}. The relevant quantum Hamiltonian reads
\bea
    H = -\frac{1}{4}\left(\partial^{ia} -mx^{ia}\right)\left(\partial_{ia} + mx_{ia}\right) + \frac{m}{2}\,\bar{\xi}_a\xi^a .\label{Hqfree}
\eea
The remaining $SU(2|1)$ generators are
\bea
    &&Q_{i} = - i\xi^a\left(\partial_{ia} -mx_{ia}\right),\qquad
    \bar{Q}_{i} = -i\bar{\xi}^a\left(\partial_{ia} + mx_{ia}\right),\nn
    &&F = -\frac{1}{2}\,\bar{\xi}_a\xi^a,\qquad I_{ik} = x^{a}_{(i}\,\partial_{k)a}\,.\label{qgenfree}
\eea
One can also define the generators of the algebra $su(2)_{\rm PG}$ for the considered case
\bea
    E_{ab} = x^{i}_{(a}\,\partial_{ib)}-\bar{\xi}_{(a}\xi_{b)}\,, \quad  \left[E_{ab},  E_{cd}\right]
    = \varepsilon_{cb} E_{ad} - \varepsilon_{ad} E_{cb}\,. \label{qext}
\eea
They can be checked to commute with all $SU(2|1)$ generators.

Since the spectrum of the Hamiltonian must be bounded from below, we define the ground state $|0\rangle$ by imposing the physical conditions
\bea
    \xi^a\,|0\rangle =0\,,\qquad \left(\partial_{ia} + mx_{ia}\right)|0\rangle =0\,.
\eea
Solving them, we obtain the ground state wave function annihilated by supercharges \eqref{qgenfree}:
\bea
    |0\rangle = e^{-\frac{m}{2}\,x^2},\qquad Q^i\,|0\rangle =\bar{Q}_i\,|0\rangle = 0\,.
\eea
All bosonic quantum states can be constructed by action of the creation operators $\nabla^{ia}:=\partial^{ia} - mx^{ia}$ on $|0\rangle$.
The bosonic state $|\ell\,;s\rangle$ is defined as
\bea
    |\ell\,;s\rangle = A_{(i_1 i_2\ldots i_s)(a_1 a_2\ldots a_s)}\nabla^{i_1 a_1}\nabla^{i_2 a_2}
    \ldots\nabla^{i_s a_s}\left(\nabla^{ia}\nabla_{ia}\right)^{\ell}|0\rangle,\label{ls}
\eea
where $A$ stand for numerical coefficients symmetric in both $SU(2)_{\rm int}$ and $SU(2)_{\rm PG}$ indices.
Here, the quantum number $s/2$ is identified with the highest weight (``isospin'') of the irreducible representation of the group $SU(2)_{\rm PG}$.
Clearly, this state has the same isospin $s/2$ with respect to $SU(2)_{\rm int}\,$.

Acting by the supercharges $\bar{Q}_i$ on the bosonic parent function $|\ell\,;s\rangle$,
we obtain the set of its fermionic descendants
\bea
    &&\bar{Q}_i\,|\ell\,;s\rangle = 2i m\,\bar{\xi}^a\left[2\ell\nabla_{i a}\,|\ell-1\,;s\rangle + s  b_{ia}\,|\ell\,;s-1\rangle\right],\nn
    &&\bar{Q}^i\bar{Q}_i\,|\ell\,;s\rangle = -8\ell\left(2\ell +s\right) m^2\left(\bar{\xi}\,\right)^2\,|\ell-1\,;s\rangle,
\eea
where $b_{ia}$ is some coefficient.
Thus, $|\ell\,;s\rangle$ extends to a super wave function $\Omega^{(\ell ; s)}$ with $\Omega^{(0 ; 0)}=|0\rangle$.
The supercharges $Q^i$ annihilate $|\ell\,;s\rangle$, i.e. $Q^i\,|\ell\,;s\rangle = 0$\,.
The spectrum of the Hamiltonian \eqref{Hqfree} is thus given by
\bea
    H\,\Omega^{(\ell ; s)} = \frac{m}{2}\left(2\ell + s\right)\Omega^{(\ell ; s)},\qquad m>0\,.\lb{Heigen}
\eea

Let us analyze the degeneracies of the superwave functions $\Omega^{(\ell ; s)}\,$, labeling the representations
of the group $SU(2)_{\rm PG}\times SU(2)_{\rm int}$ by the pair of indices  $(k,n)$.
The non-trivial wave function $\Omega^{(0;s)}$ is a superposition of $\left(s+1\right)^2$ bosonic and $s\left(s+1\right)$ fermionic states,
\bea
    |0\,;s\rangle,\qquad \bar{\xi}_a\,|0\,;s-1\rangle, \qquad s>0\,,\label{atyp}
\eea
so revealing the degeneracy $(2s +1)(s+1)$. The bosonic states correspond to the representations $\left(s/2\,, s/2\right)$
and the fermionic states to the representations $\left(s/2\,, (s-1)/2\right)$. On the other hand,
the wave functions $\Omega^{(\ell;s)}$ with $\ell > 0$ have $4\left(s+1\right)^2$-fold degeneracy, being a superposition of the following states:
\bea
    |\ell\,;s\rangle,\qquad \bar{\xi}_a \nabla^{ia}\,|\ell-1\,;s\rangle,\qquad
    \bar{\xi}_a \,|\ell\,;s-1\rangle,\qquad \left(\bar{\xi}\,\right)^2\,|\ell-1\,;s\rangle, \qquad \ell>0\,.\label{typ}
\eea
Here, the bosonic states (1st and 4th) span the representation $\left(s/2\,, s/2\right)\oplus\left(s/2\,, s/2\right)$,
while the fermionic states span the representation $\left(s/2\,,(s+1)/2\right)\oplus\left(s/2\,,(s-1)/2\right)$.

It is interesting to find out the $SU(2|1)$ representation contents of these wave functions.
For the realization \eqref{qgenfree} and \eqref{qext}, the $SU(2|1)$ Casimir operators defined in \eqref{C2}, \eqref{C3}
acquire the following concise form
\bea
    m^2 C_2=H\left(H+m\right)-\frac{m^2}{2}\,E^a_b E^b_a\,,\qquad
    m^3 C_3 = \left(H+\frac{m}{2}\right)C_2\,.\label{qC}
\eea
The last term in $C_2$ is just the Casimir of $SU(2)_{\rm PG}$ and it acts on $\Omega^{(\ell ; s)}$ as
\bea
    \frac{1}{2}\,E^a_b E^b_a\, \Omega^{(\ell ; s)} = \frac{s}{2}\left(\frac{s}{2} + 1\right)\Omega^{(\ell ; s)}.\lb{Eigen}
\eea

Now, based upon \p{qC}, \p{Eigen} and \p{Heigen}, it is easy to find the eigenvalues of $C_2$ and to cast them into the general
form given in Appendix B (eq. \eqref{Casimir}) with
\bea
    \beta =\frac{1}{2}\left(2\ell + s + 1\right),\quad \lambda = \frac{1}{2}\left(s + 1\right),\qquad\ell\neq 0\,.
\eea
The relevant quantum states form the so-called typical $SU(2|1)$ representations.
The atypical $SU(2|1)$ representations correspond to the zero eigenvalues of Casimirs, and so $\Omega^{(0 ; s)}$ belong to this subclass. For them
\bea
    \ell = 0\,,\qquad \beta = \lambda =\frac{s}{2}\,.
\eea

The degeneracy of $\Omega^{(\ell ; s)}$ can be computed as the product of the relevant dimensions of $SU(2)_{\rm PG}$
and $SU(2|1)$ representations (see Appendix \ref{appB}). The result coincides with the direct counting given above.
In the typical cases, with  $\ell>0$\,, the wave function $\Omega^{(\ell ; s)}$ has the degeneracy $4\left(s+1\right)^2$.
The typical representation always encompasses an equal number of bosons and fermions. The wave function $\Omega^{(0 ; s)}$
corresponding to the atypical case has the degeneracy $\left(2s+1\right)\left(s+1\right)$, with $\left(s+1\right)^2$ bosons and $s\left(s+1\right)$ fermions.

As instructive examples, let us consider superwave functions of the simplest atypical and typical representations of $SU(2|1)$.
The atypical superwave function $\Omega^{(0;1)}$ consists of $(1/2,1/2)$ bosonic and $(1/2,0)$ fermionic states given by
\bea
    \nabla^{i a}|0\rangle,\qquad \bar{\xi}_a |0\rangle.
\eea
The simplest typical superwave function $\Omega^{(1;0)}$ has the 4-fold degeneracy:
\bea
    \nabla^{ia}\nabla_{ia}\,|0\rangle,\qquad \bar{\xi}_a \nabla^{ia}\,|0\rangle,\qquad \bar{\xi}^2\,|0\rangle.
\eea
Here, bosonic states belong to $(0,0)\oplus(0,0)$, while fermionic states belongs to $(0,1/2)$.
\section{The `mirror' multiplet ${\bf (4,4,0)}$}
The standard multiplets ${\bf (n, 4, 4-n)}$ of the flat ${\cal N}=4, d=1$ supersymmetry have their ``mirror'' (or ``twisted'')
cousins which possess the same field contents but for which two commuting $SU(2)$ automorphism algebras of
the ${\cal N}=4, d=1$ superalgebra switch their roles. Since these automorphism algebras enter the game in the entirely symmetric way, the difference
between two mutually mirror multiplets manifests itself only in those SQM models where they are  present simultaneously.
In the $SU(2|1)$ deformed case the symmetry between the two former automorphism $SU(2)$ groups of the flat superalgebra proves to be broken:
one of these $SU(2)$ becomes $SU(2)_{\rm int}$,
while only one $U(1)$ generator $F$ from the second $SU(2)$ (which we denoted $SU^\prime(2)$) is inherited by the $SU(2|1)$ superalgebra.
So one can expect an essential difference between the $SU(2|1)$ multiplets and their possible mirror counterparts.
Here we construct the mirror version of the $SU(2|1)$ multiplet ${\bf (4, 4, 0)}$ and show that the relevant SQM models
indeed reveal a few serious distinctions from those considered in the previous sections.
In particular, in the mirror case one can define the $SU(2|1)$ invariant superfield WZ term.

Let us consider the mirror ${\bf (4,4,0)}$ multiplet \cite{DeldIv4,FIS} in the framework of the harmonic $SU(2|1)$ superspace.
The basic superfield is $\left(Y^{A}\right)^{\dagger}=\bar{Y}_{A}, A=1,2,$ satisfying
the constraints\footnote{The harmonic constraints in \p{DY} necessarily follow from the Grassmann analyticity ones as the integrability conditions for the latter,
in a close similarity with the $q^+$ multiplet.}
\bea
    \bar{\cal D}^{+} Y^{A} ={\cal D}^{+} \bar{Y}^{A} = 0\,, \quad {\cal D}^{+}Y^{A}=-\,\bar{\cal D}^{+}\bar{Y}^{A}\,,
    \quad {\cal D}^{++}Y^{A} ={\cal D}^{++}\bar{Y}^{A}=0\,.\label{DY}
\eea
With taking into account the action of the generator $\tilde{F}$ on the spinor covariant derivatives (eq. \p{Fcomm}), the
constraints \p{DY} uniquely fix the $\tilde{F}$ charges of the superfields $Y^{A}, \bar{Y}^{A} $ as
\bea
     m\tilde{F}\,\bar{Y}^{A} = -\,\frac{m}{2}\,\bar{Y}^{A},\qquad m\tilde{F}\,Y^{A} =\frac{m}{2}\,Y^{A}.\label{FY}
\eea
The  $SU^\prime(2)_{\rm PG}$ doublets $Y^{A}, \bar{Y}^A$ can be combined into a doublet $Y^{1^{\prime}A},\; Y^{2^{\prime}A}$ of $SU^{\prime}(2)$.

The constraints \p{DY} are solved in terms of the superfields defined on the analytic three-theta superspaces \eqref{AnalSubs3}:
\bea
    Y^{A}\big(\zeta_{(A)}^{(3)}\big) &=& y^{A} - \theta^{+} \psi^{iA}w^{-}_i +
\theta^{-}\psi^{iA}w^{+}_i -
    2i \,\theta^{-}\bar{\theta}^{+} \dot{y}^{A}+ 2i\,\theta^{-}\theta^{+} \dot{\bar{y}}^{A} - 2i\,\theta^{-}\theta^{+}\bar{\theta}^{+} \dot{\psi}^{iA}w^{-}_i\nn
    && +\,m\, \theta^{-}\bar{\theta}^{+} y^{A} + m\,\theta^{-}\theta^{+} \bar{y}^{A} ,\nn
    \bar{Y}^{A}\big(\bar{\zeta}_{(A)}^{(3)}\big) &=& \bar{y}^{A} - \bar{\theta}^{+} \psi^{iA}w^{-}_i + \bar{\theta}^{-} \psi^{iA}w^{+}_i
    - 2i\, \theta^{+}\bar{\theta}^{-} \dot{\bar{y}}^{A}
    + 2i \,\bar{\theta}^{+}\bar{\theta}^{-} \dot{y}^{A}- 2i\,\bar{\theta}^{-}\theta^{+}\bar{\theta}^{+} \dot{\psi}^{iA}w^{-}_i\nn
    && -\,m\, \theta^{+}\bar{\theta}^{-} \bar{y}^{A} - m\,\bar{\theta}^{+}\bar{\theta}^{-} y^{A},\label{Y}
\eea
where
\be
    \left(y^{A}\right)^\dagger = \bar{y}_{A}\,,
\qquad \left(\psi^{iA}\right)^\dagger = \psi_{iA}\,.
\ee
We observe that the field content of  $Y^{A}$ is just ${\bf (4, 4, 0)}\,$, but the $SU(2)$ assignment of the involved fields
is different from that of the previous  ${\bf (4, 4, 0)}$ multiplet. The bosonic fields are organized into a complex $SU^\prime(2)_{\rm PG}$
doublet $y^A$ which is a singlet of $SU(2)_{\rm int}$ (though it is still rotated by the $U(1)_{\rm int}$ generator
$F$); the fermionic fields are combined into doublets of both $SU^\prime(2)_{\rm PG}$ and $SU(2)_{\rm int}$,
but they are singlets of $U(1)_{\rm int}$.

According to the general transformation law \eqref{SFtr}, the superfields $Y^{A}, \bar{Y}^A$ transform as
\bea
    \delta Y^A =  - \, m \left(\bar{\theta}^{+}\epsilon^{-} - \theta^{+}\bar{\epsilon}^{-}\right)Y^A , \qquad
    \delta \bar{Y}^A = m \left(\bar{\theta}^{+}\epsilon^{-} - \theta^{+}\bar{\epsilon}^{-}\right)\bar{Y}^A .\label{trSFY}
\eea
The corresponding component field transformations are
\bea
    \delta y^{A} =-\,\epsilon_i\psi^{iA}  ,\qquad    \delta \bar{y}^A =-\,\bar{\epsilon}_i \psi^{iA},\qquad
    \delta \psi^{iA} =\bar{\epsilon}^i\left( 2i\dot{y}^{A} - m y^{A}\right)-\epsilon^i \left( 2i\dot{\bar{y}}^{A} + m\bar{y}^{A}\right).\label{trY}
\eea

\subsection{The $\sigma$-model action}
One can write the general action in terms of the function $\tilde{L}$ as
\bea
    \tilde{S}\left(Y,\bar{Y}\right) = \int dt\, \tilde{\cal L} =\int d\zeta_H \,\tilde{L}\left( Y,\bar{Y}\right).\label{Gaction-Y}
\eea
It can be checked to respect $SU(2|1)$ invariance only with the following condition
\bea
    m\tilde{F}\tilde{L}\left(Y,\bar{Y}\right) =0\qquad\Rightarrow\qquad
    m\left(y^B\partial_B -\bar{y}^B\bar{\partial}_{B}\right)\tilde{L}\left(y,\bar{y}\right) =0\,.\label{FLY}
\eea
This condition is non-trivial only for $m\neq 0$, when $F$ appears as an internal generator in the superalgebra \eqref{algebra}.
In virtue of the constraint \p{FLY},
\be
\tilde{L}(Y, \bar Y) = \tilde{L}({\cal U}^A_B), \quad {\cal U}^A_B :=  Y^A \bar{Y}_B\,.
\ee

The general component Lagrangian is
\bea
    \tilde{\cal L} &=& \left[2\,\dot{y}^{A}\dot{\bar{y}}_{A} +\frac{i}{2}\, \psi^{iA}\dot{\psi}_{iA}
    -\frac{i}{2}\,\psi^{iA}\psi_{iC}\left(\dot{y}^{C}\partial_A +\dot{\bar y}^{C}\bar{\partial}_{A}\right)
    +\frac{1}{48}\,\psi^{iA}\psi^{k}_{A}\psi_{i}^{B}\psi_{kB}\,\Delta_y\right]\tilde{G}\nn
    &&- \,im \left(\dot{y}^{A}\bar{y}_{A}-y^{A}\dot{\bar{y}}_{A}\right)\tilde{G}+ 2im \left( \dot{y}^{A}\partial_{A}\tilde{L}
    - \dot{\bar{y}}^{A} \bar{\partial}_{A}\tilde{L}\right)-m\,\psi^{iA}\psi_{i}^{B}\partial_{A}\bar{\partial}_{B}\tilde{L} \nn
    && + \,\frac{m}{4}\,\psi^{iA}\psi_{iC} \left(y^{C}\partial_A \tilde{G} - \bar{y}^{C}\bar{\partial}_{A}\tilde{G}\right)
    + \frac{m^2}{2} \,{y}^{A}{\bar{y}}_{A} \,\tilde{G}
    - m^2\left(y^{A}\partial_{A}\tilde{L} + \bar{y}^{A}\bar{\partial}_{A}\tilde{L}\right),\label{kin-Y}
\eea
where
\bea
    \tilde{G}:=\Delta_y \tilde{L} \,,\qquad \Delta_y  = -\, 2\,\epsilon^{AB}\partial_{A}\bar{\partial}_{B}\,,\qquad \partial_A
    = \frac{\partial}{\partial y^A}\,,\quad
    \bar{\partial}_B = \frac{\partial}{\partial \bar{y}^B}\,.
\eea

The simplest action is the free action
\bea
    \tilde{S}_{\rm free}\left( Y,\bar{Y}\right) = \frac{1}{4} \int d\zeta_H \, Y^{A}\bar{Y}_{A} \,.\label{action-Y}
\eea
The corresponding Lagrangian is
\bea
    \tilde{\cal L}_{\rm free}=2\,\dot{y}^{A}\dot{\bar{y}}_{A} + \frac{i}{2}\,\psi^{iA}\dot{\psi}_{iA}
    - \frac{i}{2}\,m\left(\dot{y}^{A}\bar{y}_{A} - y^{A}\dot{\bar{y}}_{A}\right).\label{LY}
\eea

In contrast to the flat $m=0$ case,  the Lagrangian \eqref{kin-Y} is not vanishing for $\tilde{G}=0$. In the latter case, the remaining
Lagrangian
\bea
    \tilde{\cal L}\left.\right|_{\tilde{G}=0} = 2im \left( \dot{y}^{A}\partial_{A}\tilde{L}
    - \dot{\bar{y}}^{A} \bar{\partial}_{A}\tilde{L}\right)-m\,\psi^{iA}\psi_{i}^{B}\partial_{A}\bar{\partial}_{B}\tilde{L}
    - m^2\left(y^{A}\partial_{A}\tilde{L} + \bar{y}^{A}\bar{\partial}_{A}\tilde{L}\right)\label{kin-Y1}
\eea
is $SU(2|1)$ invariant, and it can be considered as a WZ Lagrangian vanishing at $m=0$. The condition $\tilde{G}=0$
is equivalent to the dim 4 Laplace equation for $\tilde{L}$
\be
\Delta_y \tilde{L} = 0\,.\lb{lapl}
\ee
The simplest non-trivial solution of \p{lapl} is
\be
\tilde{L} = c_{(AB)}Y^A\bar Y^B\,, \lb{SImpl}
\ee
where $c_{(AB)}$ is an arbitrary constant triplet which breaks $SU(2)_{\rm PG}$ symmetry. Another, $SU(2)_{\rm PG}$ invariant  solution is
\be
\tilde{L} \sim \frac{1}{y^2}\,, \quad y^2 =  y^A \bar{y}_A\,.
\ee
The singularity at $y^A =0$ can be avoided  by assuming that the fields $y^A$ have a non-trivial constant vacuum background.  Other
possible solutions of \p{lapl} are famous multi-center solutions breaking $SU(2)_{\rm PG}$ invariance.

\subsection{Wess-Zumino action}
For the mirror ${\bf (4, 4, 0)}$ multiplet, the WZ term can be obtained not only as a special limit of the $SU(2|1)$ invariant $\sigma$-model term,
but can be also constructed independently. The superfield WZ action can be written as an integral over the analytic superspace
\bea
    \tilde{S}_{\rm WZ}\left(Y,\bar{Y}\right)
    =-\,\gamma \int  d\zeta_{(A)}^{--}\,\left(\bar{\theta}^{+}\bar{\cal D}^{+} + \theta^{+}{\cal D}^{+}\right)f\left(Y,\bar{Y}\right).\label{WZact}
\eea
Since we integrate over the analytic superspace, we need to impose the analyticity condition \eqref{A-cond} on the superfield Lagrangian.
It gives the condition
\bea
    \bar{\cal D}^{+}{\cal D}^{+}f\left(Y,\bar{Y}\right)=0 \quad\Rightarrow\quad \Delta_y f = 0\,.\label{self}
\eea
One can define the background 4-vector $\tilde{\cal A}_{i^\prime A}$ as
\bea
    \tilde{\cal A}_{1^{\prime}B}= i \partial_{B} f,\quad \tilde{\cal A}_{2^{\prime}B}=
    -\, i \bar{\partial}_{B} f,\quad y^{A}=y^{1^{\prime}A},\quad y^{2^{\prime}A}=\bar{y}^{A}.\label{4v-A}
\eea
Then eq. \p{self} implies the self-duality condition for $\tilde{\cal A}_{i^\prime A}$
\bea
    \partial_{i^\prime A}\tilde{\cal A}_{k^\prime B}-\partial_{k^\prime B}\tilde{\cal A}_{i^\prime A}= \epsilon_{i^\prime k^\prime}
    \tilde{\cal B}_{(AB)}\,, \quad \tilde{\cal B}_{(AB)} = -2i\partial_{(A}\bar\partial_{B)} f\,,
\eea
as well as the  transversal gauge condition
\bea
    \partial_{i^\prime A}\tilde{\cal A}^{i^\prime A}=0\,.
\eea

In addition, the requirement of $SU(2|1)$ invariance gives rise  to a new constraint for $f$ at $m\neq 0$ (cf. \p{FLY}):
\bea
    m\tilde{F}f\left(Y,\bar{Y}\right) =0\qquad\Rightarrow\qquad m\left(y^B\partial_B
    -\bar{y}^B\bar{\partial}_{B}\right)f\left(y,\bar{y}\right)  =  0\,.\label{U11}
\eea
This condition amounts to the invariance of \eqref{WZact} under the internal $U(1)$ symmetry
\be
f\left(Y^A,\bar{Y}_B\right) = f({\cal U}^A_B)\,, \quad {\cal U}^A_B :=Y^A\bar{Y}_B\,.\label{UAB}
\ee

In the limit $m=0$, the matrix generator $\tilde{F}$ becomes an external automorphism generator and
the condition \eqref{U11} is satisfied trivially, without imposing any constraints on $f\left(Y_A,\bar{Y}^B\right)$.

The component Lagrangian corresponding to the action \p{WZact} reads
\bea
    \tilde{\cal L}_{\rm WZ} = 2\gamma\, \Big\lbrace i \left(\dot{y}^{A}\,\partial_{A}f - \dot{\bar{y}}^{A}\,\bar{\partial}_{A}f\right)
    - \frac{m}{2}\left(y^{A}\,\partial_{A}f + \bar{y}^{A}\bar{\partial}_{A}f\right)
    - \frac{1}{2}\,\psi^{iA}\psi^{B}_{i}\,\partial_{A}\bar{\partial}_{B}f\,\Big\rbrace .\label{WZ-Y}
\eea
Employing the conditions \eqref{self}, \eqref{U11}, one can directly check that this Lagrangian
is invariant under the supersymmetry transformations \eqref{trY}. The first term in \p{WZ-Y} can be concisely
rewritten through the external gauge field as
\bea
 i(\dot{y}^{A}\,\partial_{A}f - \dot{\bar{y}}^{A}\,\bar{\partial}_{A}) = \dot{y}^{i'A}{\cal A}_{i'A}\,. \label{WZ-Y1}
\eea
Note that the $SU^\prime(2)$ symmetry is broken in the full Lagrangian \p{WZ-Y}.

Clearly, the Lagrangian \eqref{WZ-Y} can be identified with the Lagrangian \eqref{kin-Y1}, where
\bea
    \tilde{G} = \Delta_y \tilde{L}\left(y,\bar{y}\right)  =0\,,\qquad\tilde{L}\equiv f,\qquad m\sim \gamma\,.
\eea
The constraint \eqref{U11} amounts to the condition \eqref{FLY}.

\subsection{Hamiltonian formalism and quantum supercharges}
We start with the total Lagrangian $\tilde{\cal L}+\tilde{\cal L}_{\rm WZ}$\,.
The corresponding canonical Hamiltonian reads
\bea
    H&=& -\,\frac{1}{2}\,\epsilon^{AB}\tilde{G}^{-1}\left[p_{A}-2i\left(m\,\partial_A \tilde{L}+\gamma\,\partial_A f\right)
    + \frac{i}{2}\,\psi^{kD}\psi_{kA}\,\partial_{D}\tilde{G}\right]\times\nn
    &&\left[\bar{p}_B + 2i\left(m\,\bar{\partial}_B\tilde{L}+\gamma\,\bar{\partial}_B f\right)
    + \frac{i}{2}\,\psi^{iC}\psi_{iB}\,\bar{\partial}_{C}\tilde{G}\right]-\frac{1}{48}\,\psi^{iA}\psi^{k}_{A}\psi_{i}^{B}\psi_{kB}\,\Delta_y\tilde{G}\nn
    &&+\,\psi^{iA}\psi_{i}^{B}\left(m\,\partial_{A}\bar{\partial}_{B}\tilde{L}+\gamma\,\partial_{A}\bar{\partial}_{B}f\right)-\frac{i}{2}\,m\left(y^A p_A
    - \bar{y}^A\bar{p}_A\right).
\eea
The relevant supercharges and the rest of bosonic generators are given by
\bea
    &&Q^i = \psi^{iA}\left[p_{A}-2i\left(m\,\partial_A \tilde{L}+\gamma\,\partial_A f\right) +\frac{i}{6}\,\psi^{kC}\psi_{kA}\,\partial_{C}\tilde{G}\right],\nn
    &&\bar{Q}_i = \psi_{i}^{A}\left[\bar{p}_{A} + 2i\left(m\,\bar{\partial}_A\tilde{L}+\gamma\,\bar{\partial}_A f\right)
    + \frac{i}{6}\,\psi^{kC}\psi_{kA}\,\bar{\partial}_{C}\tilde{G}\right],\lb{Qclass2} \\
    &&F= -\frac{i}{2}\left(y^A p_A - \bar{y}^A\bar{p}_A\right),\qquad I_{ik} = \frac{1}{2}\,\psi^{A}_{(i}\psi_{k)A}\,\tilde{G}\,.
\eea

We impose the Poisson brackets and the Dirac brackets as
\bea
    &&\lbrace p_{A}, y^{B}\rbrace = -\,\delta_A^B\,,\qquad\lbrace \bar{p}_{A}, \bar{y}^{B}\rbrace = -\,\delta_A^B\,,\qquad
    \lbrace \psi^{iA},\psi_{kB}\rbrace = -\,i\,\tilde{G}^{-1}\delta_B^A\,\delta^i_k\,,\nn
    &&\lbrace p_{A}, \psi_{kB}\rbrace = \frac{1}{2}\,\psi_{kB}\,\tilde{G}^{-1}\partial_A\,\tilde{G}\,,\qquad
    \lbrace \bar{p}_{A}, \psi_{kB}\rbrace = \frac{1}{2}\,\psi_{kB}\,\tilde{G}^{-1}\bar{\partial}_A\tilde{G}\,.
\eea
After redefining
\bea
    \psi^{iA} =\tilde{G}^{-\frac{1}{2}}\xi^{iA},
\eea
the brackets are simplified to
\bea
    \lbrace p_{A}, y^{B}\rbrace = -\,\delta_A^B\,,\qquad\lbrace \bar{p}_{A}, \bar{y}^{B}\rbrace = -\,\delta_A^B\,,\qquad
    \lbrace \xi^{iA},\xi_{kB}\rbrace = -\,i\,\delta_B^A\,\delta^i_k\,.
\eea
We quantize these brackets in the standard way
\bea
    &&p_{A}=-\,i\partial_A\,,\qquad \bar{p}_{A}=-\,i\bar{\partial}_{A}\,,\qquad \xi_{iA}=\partial/\partial\xi^{iA}\,,\nn
    &&\left[p_{A}, y^{B}\right] = -\,i\,\delta_A^B\,,\qquad\left[\bar{p}_{A}, \bar{y}^{B}\right] =
    -\,i\,\delta_A^B\,,\qquad\lbrace \xi^{iA},\xi_{kB}\rbrace = \delta_B^A\,\delta^i_k\,.
\eea

The quantum supercharges obtained from the classical ones  by the general procedure of ref. \cite{How} are given by
\bea
    &&Q^i_{\rm (cov)} = -i\tilde{G}^{-\frac{1}{2}}\xi^{iA}\left[\partial_{A} +2 \left(m\,\partial_A\tilde{L}+\gamma\,\partial_A f\right)
    -\frac{1}{6}\left(\xi^{kC}\xi_{kA}-4\delta^{C}_{A}\,\right)\tilde{G}^{-1}\partial_{C}\tilde{G}\right],\nn
    &&\bar{Q}_{{\rm (cov)}i} = -i\tilde{G}^{-\frac{1}{2}}\xi_{i}^{A}\left[\bar{\partial}_{A}
    - 2\left(m\,\bar{\partial}_A\tilde{L}+\gamma\,\bar{\partial}_A f\right)
    - \frac{1}{6}\left(\xi^{kC}\xi_{kA}-4\delta^{C}_{A}\,\right)\tilde{G}^{-1}\bar{\partial}_{C}\tilde{G}\right].\lb{Qquant2}
\eea
Their anticommutator yields the quantum Hamiltonian
\bea
    H_{\rm (cov)}&=& \,\frac{1}{4}\,\epsilon^{AB}\tilde{G}^{-1}\bigg\lbrace\left[\partial_{A}
    +2\left(m\,\partial_A \tilde{L}+\gamma\,\partial_A f\right)
    + \frac{1}{2}\,\xi_{kA}\xi^{kD}\,\tilde{G}^{-1}\partial_{D}\tilde{G}\right]\times\nn
    &&\left[\bar{\partial}_B - 2\left(m\,\bar{\partial}_B\tilde{L}+\gamma\,\bar{\partial}_B f\right)
    + \frac{1}{2}\,\xi_{iB}\xi^{iC}\,\tilde{G}^{-1}\bar{\partial}_{C}\tilde{G}\right]\bigg\rbrace\nn
    &&+\,\frac{1}{48}\left[\xi^{iA}\xi_{i}^{B}\xi^{k}_{B}\xi_{kA}+2\right]\tilde{G}^{-2}\Delta_y\tilde{G}
    +\xi^{iA}\xi_{i}^{B}\,\tilde{G}^{-1}\left(m\,\partial_{A}\bar{\partial}_{B}\tilde{L}+\gamma\,\partial_{A}\bar{\partial}_{B}f\right)\nn
    &&-\,\frac{m}{2}\left(y^A \partial_A - \bar{y}^A\bar{\partial}_A\right) -\frac{m}{2},
\eea
as well as the remaining  bosonic generators:
\bea
F= -\frac{1}{2}\left(y^A \partial_A - \bar{y}^A\bar{\partial}_A\right),\qquad I_{ik} = \frac{1}{2}\,\xi^{A}_{(i}\xi_{k)A}\,.
\eea

In analogy with \eqref{W-rep}, we can represent the supercharges as
\bea
    Q_{{\rm (cov)}i} = e^{-2m\tilde{L}}Q_{{\rm (cov)}i}^{\left(m=0\right)}e^{2m\tilde{L}},\qquad
    \bar{Q}_{{\rm (cov)}i} = e^{2m\tilde{L}}\bar{Q}_{{\rm (cov)}i}^{\left(m=0\right)}e^{-2m\tilde{L}},
\eea
thereby relating them to the particular case of the undeformed supercharges of the ``flat'' mirror ${\bf (4, 4, 0)}$ multiplet.

\subsection{The free model}
As an instructive example, here we consider the free model with
$\tilde{G}=1$ and the simplest WZ term \eqref{WZ-Y} in which the
function $f$ has been  chosen as (recall \p{SImpl}) \bea
    f=\frac{1}{4}\, c_{AB}\,y^{A}\bar{y}^{B},\qquad c_{AB}=c_{BA}\,,\qquad \overline{\left(c_{AB}\right)} = -\,c^{BA}.
\eea
The corresponding total component Lagrangian reads:
\bea
\tilde{{\cal L}}_{\rm free} &=&2\,\dot{y}^{A}\dot{\bar{y}}_{A} + \frac{i}{2}\,\xi^{iA}\dot{\xi}_{iA} - \frac{i}{2}\,m\left(\dot{y}^{A}\bar{y}_{A} - y^{A}\dot{\bar{y}}_{A}\right)\nn
    && +\, \frac{\gamma}{2}\, c_{AB}\, \Big\lbrace i\left(\dot{y}^{A}\bar{y}^B - \dot{\bar{y}}^{A}y^B\right)
    - \frac{m}{2}\left(y^{A}\bar{y}^B + \bar{y}^{A}y^B\right) - \frac{1}{2}\,\xi^{iA}\xi^{B}_{i}\Big\rbrace .\label{LYfree}
\eea
One can choose the $SU(2)_{\rm PG}$ frame  in which the triplet $c_{AB}$ is reduced to
\bea
c_{12}=c_{21}=1\,, \qquad c_{11}=c_{22}=0\,.\lb{frame}
\eea

For further use, we define the operators
\bea
    \nabla^{\pm}_A = \partial_{A} \pm \frac{1}{2}\left(m\,\epsilon_{AC} + \gamma\,c_{AC}\right)\bar{y}^C ,\qquad
    \bar{\nabla}^{\pm}_B=\bar{\partial}_B \pm \frac{1}{2}\left(m\,\epsilon_{BD} - \gamma\,c_{BD}\right)y^D ,\label{nablagamma}
\eea
which form the following algebra
\bea
    \left[\nabla^{\pm}_A,  \bar{\nabla}^{\pm}_B\right] = \mp\left(m\,\epsilon_{AB} + \gamma\,c_{AB}\right).
\eea
The quantum Hamiltonian reads
\bea
    H = \frac{1}{2}\,\epsilon^{AB}\bar{\nabla}^{+}_B\nabla^{+}_{A}
    +\frac{\gamma}{4}\,c_{AB}\,\xi^{iA}\xi_{i}^{B}  - \frac{m}{2}\left(y^A \partial_A - \bar{y}^A\bar{\partial}_A\right) + \frac{m}{2}\,,\label{Hgamma}
\eea
and the remaining $SU(2|1)$ generators are written as
\bea
    &&Q^i = -\,i\,\xi^{iA}\nabla^{+}_{A}\,,\qquad \bar{Q}_{i} = -\,i\,\xi_{i}^{B}\bar{\nabla}^{+}_B\,,\nn
    &&F= -\frac{1}{2}\left(y^A \partial_A - \bar{y}^A\bar{\partial}_A\right),\qquad I_{ik} = \frac{1}{2}\,\xi^{A}_{(i}\xi_{k)A}\,.\label{gengamma}
\eea

We construct the bosonic wave functions in terms of the operators $\nabla^{\pm}_{A}$, $\bar{\nabla}^{\pm}_A$ satisfying
\bea
    &&\left[H,\nabla^{+}_{A}\right] = \frac{\gamma}{2}\,c^{B}_{A}\,\nabla^{+}_{B}\,,\qquad
    \left[H,\bar{\nabla}^{+}_{A}\right] = \frac{\gamma}{2}\,c^{B}_{A}\,\bar{\nabla}^{+}_{B}\,,\nn
    &&\left[H,\nabla^{-}_{A}\right] = \frac{m}{2}\,\nabla^{-}_{A}\,,\qquad
    \left[H,\bar{\nabla}^{-}_{A}\right] = -\frac{m}{2}\,\bar{\nabla}^{-}_{A}\,.
\eea
Imposing the following physical conditions
\bea
    \bar{\nabla}^{+}_1\,|0\rangle =\nabla^{+}_{1}\,|0\rangle =\bar{\nabla}^{-}_A\,|0\rangle =0\,,\qquad\xi^{i2}\,|0\rangle =0\,,\label{Ycond}
\eea
we make sure that the spectrum of the Hamiltonian \eqref{Hgamma} is bounded from below for $\gamma>0$ and $m>0$\,,
so that the ground state $|0\rangle$ is the lowest level. From these conditions, we find that $m=\gamma$ and the ground state wave function is obtained as
\bea
    |0\rangle = e^{-m y^1\bar{y}_1}.
\eea
The conditions \eqref{Ycond} firmly imply that the ground state $|0\rangle$ is annihilated by both supercharges defined in \eqref{gengamma}:
\bea
    Q^i\,|0\rangle = \bar{Q}_i\,|0\rangle = 0\,.
\eea
The ground state acquires the minimal energy value $E=0$ and the Casimir $C_2$ is vanishing on it.

For the choice $\gamma =m$ the Hamiltonian \eqref{Hgamma} takes the form
\bea
    H = \frac{1}{2}\,\epsilon^{AB}\bar{\nabla}^{+}_B\nabla^{+}_{A}
    +\frac{m}{2}\,\xi^{i1}\xi_{i}^{2} - \frac{m}{2}\left(y^A \partial_A - \bar{y}^A\bar{\partial}_A\right),\label{Hgamma2}
\eea
where
\bea
    &&\nabla^{\pm}_1 = \partial_{1} \pm m\,\bar{y}_1\,,\qquad\bar{\nabla}^{\pm}_1=\bar{\partial}_1 \,,\qquad
    \nabla^{\pm}_2=\partial_{2}\,,\qquad   \bar{\nabla}^{\pm}_2=\bar{\partial}_2 \pm m\,y_2 \,.
\eea
The generators \eqref{gengamma} can also be written through these operators.
The higher-order bosonic states must be constructed in terms of the operators $\nabla^{-}_{1}$, $\bar{\nabla}^{+}_2$
the commutators of which with the Hamiltonian are as follows
\bea
    \left[H,\bar{\nabla}^{+}_2\right]=\frac{m}{2}\,\bar{\nabla}^{+}_2\,,\qquad \left[H,\nabla^{-}_1\right]= \frac{m}{2}\,\nabla^{-}_1\,.
\eea
The bosonic state $|\ell\,;n\rangle$ is defined as
\bea
    |\ell\,;n\rangle = \left(\nabla^{-}_{1}\right)^n \left(\bar{\nabla}^{+}_2\right)^{\ell}|0\rangle.
\eea
Superwave function $\Omega^{(\ell ; n)}$ is obtained as a sum of the relevant fermionic descendants of $|\ell\,;n\rangle$
produced by action of the supercharges as
\bea
    &&Q^i\,|\ell\,;n\rangle = 2 i \ell m\,\xi^{i1}\,|\ell - 1\,;n\rangle,\quad
    Q^2\,|\ell\,;n\rangle = 4\ell\left(\ell - 1\right)m^2\,\xi^{i1}\xi^{1}_{i}\,|\ell - 2\,;n\rangle,\nn
    &&\bar{Q}_i\,|\ell\,;n\rangle = 0\,.
\eea
Here, $\Omega^{(0 ; 0)}=|0\rangle$. One can see that the functions $\Omega^{(0 ; n)}$ and $\Omega^{(1 ; n)}$ form singlet and triplet states, respectively.

Then, the eigenvalues of \eqref{Hgamma} are
\bea
    H\,\Omega^{(\ell ; n)} = \frac{m}{2}\left(\ell + n\right)\Omega^{(\ell ; n)},\qquad m>0\,.
\eea
The eigenvalues \eqref{Casimir} of Casimir operators are given by
\bea
    &&\beta = \frac{\ell}{2}\,,\qquad \lambda = \frac{1}{2}\,,\quad{\rm for}\quad\ell\neq 0\,,\nn
    &&\beta = \lambda = 0\,,\quad{\rm for}\quad\ell = 0\,.
\eea
Casimir operators take zero eigenvalues for $\ell = 0$ and $\ell = 1$\,.
Hence, the functions $\Omega^{(0 ; n)}$ and $\Omega^{(1 ; n)}$ correspond to atypical representations of $SU(2|1)$ with non-equal
number of bosonic and fermionic states. The functions $\Omega^{(\ell ; n)}$ with $\ell\geqslant 2$ correspond to the typical 4-fold representations of $SU(2|1)$.
The same degeneracies with respect to $\ell$ were observed in \cite{DSQM}.

In fact, this quantum free model can be identified with one of the  $SU(2|1)$ invariant models on a complex plane considered in \cite{DSQM}.

One can exclude the operators $\nabla^{\pm}_2$ and $\bar{\nabla}^{\pm}_1$ from the sets \eqref{Hgamma2} and \eqref{gengamma},
since they annihilate the ground state $|0\rangle$ and all other states.
Then, if we redefine the system as
\bea
    &&\xi^{i1}\rightarrow \eta^i,\qquad \xi^{i2}\rightarrow \bar{\eta}^i,\qquad y^1\rightarrow z\,,\qquad \bar{y}_1\rightarrow \bar{z}\,,\nn
    &&m\rightarrow m/2\,,\quad H\rightarrow H/2\,,\quad \sqrt{2}\,Q^i\rightarrow Q^i,\quad \sqrt{2}\,\bar{Q}_i\rightarrow\bar{Q}_i\,, \quad n\rightarrow n+\ell\,,
\eea
we obtain exactly the space of states of the model on a complex plane with $\kappa=1/4$ constructed by us in \cite{DSQM}
for the $SU(2|1)$ multiplet ${\bf (2,4,2)}$.

This correspondence can be understood from the viewpoint of the Hamiltonian reduction with respect to some shift isometry of the Lagrangian \p{LYfree}
with $\gamma =m$. Indeed, in the frame \p{frame} it can be written as
\bea
    \tilde{{\cal L}}_{\rm free}=2\,\dot{y}^{1}\dot{\bar{y}}_{1} + \frac{i}{2}\,\xi^{iA}\dot{\xi}_{iA} - \frac{m}{2}\,\xi^{i1}\xi^{2}_{i} - \frac{m^2}{2}\,y^{1}\bar{y}_1 +
    2\left(\dot{y}^{\,2} - \frac{i}{2}\, {m}\, y^{\,2}\right)\left(\dot{\bar y}_{\,2} +\frac{i}{2}\, {m}\, \bar y_{\,2}\right),
\eea
and it is invariant under the transformations $\delta y^{\,2} = \beta\, e^{\frac{i}2 m t}\,, \; \delta \bar y_{\,2} = \bar\beta\, e^{-\frac{i}{2} m t}\,$,
$\beta$ being a constant complex parameter.

Finally, let us briefly discuss the interesting case with $\gamma=0\,, m\neq 0$\,. In this case the Hamiltonian becomes purely bosonic
and commutes with $\xi^{iA}$ and $\nabla^{+}_{A}$\,, $\bar{\nabla}^{+}_{B}$\,, which can be treated as the generators of the ``magnetic supertranslations'' in
the target space (the Lagrangian \p{LYfree} at $\gamma =0$ is invariant under  the independent shifts of the variables $\xi^{iA}$ and $y^A$).
We rewrite the Hamiltonian \eqref{Hgamma} as \bea
    H = \frac{1}{2}\,\epsilon^{AB}\nabla^{-}_{A}\bar{\nabla}^{-}_B + \frac{m}{2}\,.
\eea
By imposing the physical condition $\bar{\nabla}^{-}_A\,\Psi_0 = 0$\,, we define the superwave function $\Psi_0$ with the lowest energy $m/2$\,:
\bea
    \Psi_0 =\left[ g_0\left(y^A\right)+g_i\left(y^A\right)\xi^{i1}+g_1\left(y^A\right)\xi^{i1}\xi^{1}_{i}\right]|0\rangle ,
    \qquad |0\rangle = e^{-\frac{m}{2}\,y^A\bar{y}_A},\qquad \xi^{i2}\,|0\rangle=0\,.
\eea
Here $g_0\,, g_i\,, g_1$ are arbitrary holomorphic polynomials of $y^A$ which all can be generated through action of $\bar{\nabla}^{+}_B$ on $|0\rangle $,
taking into account that $\nabla^{+}_B\, |0\rangle =0$\,. The infinite degeneracy of the ground state $\Psi_0$ is just due to the above mentioned symmetry of \p{LYfree}
at $\gamma =0$ under the ``magnetic supertranslation'' group.

Casimir operators \eqref{C2}, \eqref{C3} take zero eigenvalues only on the following states:
\bea
    \Omega_0 = \left(\bar{\nabla}^{+}_2 - \frac{1}{2}\,\xi^{i1}\xi^{1}_{i}\bar{\nabla}^{+}_1\right)|0\rangle ,\qquad \Omega^i=\xi^{i1} |0\rangle .
\eea
These one bosonic and two fermionic states form the fundamental  atypical representation of $SU(2|1)$,
and the supercharges \eqref{gengamma} act on them as follows
\bea
    &&Q^i\,\Omega_0 = im\,\Omega^i,\qquad \bar{Q}_i\,\Omega_0 = 0\,,\nn
    &&Q^k\,\Omega^i = 0\,,\qquad \bar{Q}_k\,\Omega^i = -\,i \delta^i_k\,\Omega_0\,.
\eea
All other states correspond to the $4$-fold typical representations of $SU(2|1)$. For instance, the state $|0\rangle $ is a component of the following $SU(2|1)$
supermultiplet
\bea
|0\rangle\,, \quad \xi^1_i\bar\nabla^+_1|0\rangle\,, \quad \left(\frac12\, \xi^{1i}\xi^1_i\bar\nabla^+_1\bar\nabla^+_1 - \bar\nabla^+_1\bar\nabla^+_2 \right)|0\rangle\,.
\eea
Thus the ground state superwave function $\Psi_0 $ is represented as an infinite sum of non-singlet states of $SU(2|1)$.
In other words, among states from which $\Psi_0 $ is composed there is no state which would be simultaneously
annihilated by both supercharges, $Q^i$ and $\bar{Q}_i$. So $SU(2|1)$ supersymmetry is spontaneously broken at $\gamma =0\,$, in contrast
to the case of $\gamma = m\neq 0\,$. The option with $\gamma \neq m$ also features the spontaneous breaking of $SU(2|1)\,$.

\section{Superconformal models}\label{section 5}
In this section we consider trigonometric superconformal models built on the  ${\bf (4,4,0)}$ $SU(2|1)$ superfields.
Their construction and basic features closely follow the pattern of superconformal models for the $SU(2|1)$ multiplets
${\bf (1, 4, 3)}$ and ${\bf (2, 4, 2)}$ discussed in \cite{ISTconf}.

\subsection{Generalities}
The most general ${\cal N}=4, d=1$ superconformal algebra is $D(2,1;\alpha)$ \cite{superc,Sorba}, $\alpha$ being a real parameter.
The proper embedding of the superalgebra $su(2|1)$ into $D(2,1;\alpha)$ is achieved through
the following redefinition of the $U(1)$ generator $\tilde{H}$ in \eqref{algebra-1}:
\bea
    \tilde{H} = {\cal H} + \left(1+\alpha\right) \mu F,\qquad m = -\,\alpha\mu\,. \label{H-cal}
\eea
Correspondingly, the $su(2|1)$ superalgebra \eqref{algebra-1} is redefined as
\bea
    &&\lbrace Q^{i}, \bar{Q}_{j}\rbrace = -\,2\alpha\mu\, I^i_j + 2\delta^i_j [{\cal H} + \left(1+\alpha\right)\mu F]\, ,\nn
    &&\left[I^i_j, \bar{Q}_{l}\right] = \frac{1}{2}\delta^i_j\bar{Q}_{l}-\delta^i_l\bar{Q}_{j}\, ,\qquad \left[I^i_j, Q^{k}\right]
    = \delta^k_j Q^{i} - \frac{1}{2}\delta^i_j Q^{k},\nn
    &&\left[F, \bar{Q}_{l}\right]=-\frac{1}{2}\,\bar{Q}_{l}\,,\qquad \left[F, Q^{k}\right]=\frac{1}{2}\,Q^{k},\nn
    &&\left[{\cal H}, \bar{Q}_{l}\right]= \frac{\mu}{2}\,\bar{Q}_{l}\,,\qquad \left[{\cal H}, Q^{k}\right]= -\frac{\mu}{2}\, Q^{k}.  \label{alpha}
\eea
The particular choice $\alpha=-1$ in \p{alpha} yields
\eqref{algebra-1} at $m=\mu\,$, with $F$ being decoupled and becoming an external
automorphism generator.  At $\alpha = 0$, the
superalgebra \eqref{alpha} converts into some $U(1)$-deformed
``Poincar\'e'' ${\cal N}=4, d=1$ superalgebra,
\bea
\lbrace Q^{i}, \bar{Q}_{j}\rbrace = 2\delta^i_j \left({\cal H} + \mu\,F\right)\,, \label{alpha1}
\eea
with $I^i_k$ becoming some external $SU(2)$ automorphism generators
(the remaining non-zero commutation relations are the same as in \p{alpha}).

The basic virtue  of rewriting the superalgebra $su(2|1)$ in the form
\eqref{alpha} is that the closure of \eqref{alpha} with its $-\mu$
counterpart in the properly defined basis of the $SU(2|1)$
superspace is just ${\cal N}=4$ superconformal algebra
$D(2,1;\alpha)$ in the so called trigonometric realization
\cite{ISTconf}. This remarkable property implies the simple
criterion for one or another $SU(2|1)$ invariant action to be
superconformal: It should be an even function of the parameter
$\mu\,$. The same property applies to the case $\alpha =0\,$ too. Once again, $D(2,1;0)$ is contained in a closure of \p{alpha1} with its $-\mu$ counterpart.
The generator ${\cal H}$ in \p{alpha}, \p{alpha1} is embedded into the $d=1$ conformal algebra $so(2,1)$ in
the standard basis\footnote{This is the basis, in which $[\hat{D}, \hat{H}] =-i\hat{H}\,, \; [\hat{D}, \hat{K}] =i\hat{K}\,, \; [\hat{H}, \hat{K}] = 2i\hat{D}\,$.}
as \cite{ISTconf}
\be
{\cal H} = \hat{H} + \frac{\mu^2}{4}\hat{K}\,.
\ee

To ensure the analogous property in application to the harmonic and analytic $SU(2|1)$ superspaces, we need to appropriately modify
the original definitions of these superspaces. We again pass to the notations \eqref{sub+} and rewrite the superalgebra \eqref{alpha}
in terms of the generators $\left\lbrace Q^{\pm},\bar{Q}^{\pm},I^{++}, I^{--}, I^{0}\right\rbrace\,$.
Then we extend this superalgebra by the $SU(2)_{\rm ext}$ generators $\{T^0,T^{++},T^{--}\}\,$ and define the $\alpha \neq 0$
analogs of the harmonic $SU(2|1)$ superspace \eqref{HB} in the analytic basis as the coset superspaces
\bea
    &&\frac{\{{\cal H}, Q^\pm, \bar Q^\pm, F, I^{\pm\pm}, I^0,  T^{\pm\pm}, T^0\}}{\{F, I^{++}, I^0, I^{--} - T^{--}, T^0\}}\,,\qquad \alpha \neq 0\,,-1\,,\nn
    &&\frac{\{{\cal H}, Q^\pm, \bar Q^\pm, I^{\pm\pm}, I^0,  T^{\pm\pm}, T^0\}}{\{I^{++}, I^0, I^{--} - T^{--}, T^0\}}\,,\qquad\alpha = -1\,.\label{coset-1}
\eea
The relevant $\epsilon, \bar\epsilon$ coordinate transformations are obtained from \p{HBtr} via the substitution $m \rightarrow -\alpha\mu$
and the redefinition \eqref{subs2} of the Grassmann coordinates, such that the generator ${\cal H}$ becomes the purely time-translation one,
${\cal H} =i \partial/\partial t_{(A)}$ \cite{ISTconf}.
These transformations together with their $-\mu$ counterparts produce, as a closure, some realization of
$D(2,1;\alpha)$ on the harmonic superspace coordinates.  This realization preserves the $SU(2|1)$ harmonic analyticity,
but only in the $\mu = 0$ limit it reproduces the realization given in \cite{IvLe}. The main difference between the  $\mu\neq 0$
and $\mu=0$ options is that the former yields the {\it trigonometric} realization of the $d=1$ conformal $SO(2,1)$ transformations \cite{Trigonom}, while the latter
gives rise to their standard {\it parabolic} realization.

For the special case $\alpha=0$, the appropriate coset superspace is defined as
\bea
    \frac{\{{\cal H}, Q^\pm, \bar Q^\pm ,F, T^{\pm\pm}, T^0\}}{\{ F, T^0\}}\,,\qquad\alpha = 0\,.\label{coset-2}
\eea
Despite the fact that  the coordinate transformations \eqref{HBtr} at $\alpha=0$ take the form corresponding
to the standard flat harmonic ${\cal N} = 4, d=1$ superspace \cite{IvLe} (because $m = -\alpha\mu$), superconformal transformations appearing as
a closure of these $\alpha=0$ transformations (in some $\mu$-dependent superspace basis) with their $-\mu$ counterparts  still constitute the  trigonometric realization
of $D(2,1;0)$. Note that  the harmonic variables $w^\pm_i$ in the $\alpha=0$ case do not transform  under the $\epsilon, \bar\epsilon$
transformations at all (as well as under the full set of the $D(2,1;0)$ transformations, including those of the relevant $SU(2)_{\rm int}$, which act
only on the Grassmann coordinates).

For the generic $\alpha$\,, the trigonometric superconformal transformations of the deformed ${\cal N}=4, d=1$ superspace coordinates
were given in \cite{ISTconf}. In order to find the trigonometric realization in the analytic basis of harmonic superspace,
one should use the relations \eqref{subs} with $m=-\alpha\mu$ or the relations \eqref{subs1}.
The basic relations of the CR structure \eqref{CRstruct} are rewritten as
\bea
    \{{\cal D}^{+}, \bar{{\cal D}}^{+}\}= 2\alpha\mu\,{\cal D}^{++},\qquad
\left[{\cal D}^{++},  {\cal D}^+ \right] = \left[{\cal D}^{++},  \bar{{\cal D}}^+ \right] = 0\,,\lb{CRstruct1}
\eea
where the covariant derivatives ${\cal D}^{+}$, $\bar{{\cal D}}^{+}$ and ${\cal D}^{++}$ (in the original coordinates)
are given by the following expressions valid for any $\alpha$:
\bea
    {\cal D}^{+} &=& e^{-\frac{i}{2}\mu  t_{(A)}}\left[\frac{\partial}{\partial \theta^-} -\alpha\mu\,\bar{\theta}^{-}{\cal D}^{++}\right],\quad
    \bar{{\cal D}}^{+} = e^{\frac{i}{2}\mu t_{(A)}}\left[-\frac{\partial}{\partial \bar\theta^-} +\alpha\mu\,\theta^{-}{\cal D}^{++}\right],\nn
    {\cal D}^{++} &=& \left(1 -\alpha\mu\,\bar{\theta}^{+}\theta^{-} + \alpha\mu\, \bar{\theta}^{-}\theta^{+}\right)^{-1} \partial^{++}
    + 2i\, \theta^+\bar\theta^{+}\partial_{(A)} +2\left(1+\alpha\right)\mu\, \theta^{+}\bar\theta^{+}\tilde{F} \nn
    &&+\, \theta^{+}\frac{\partial}{\partial \theta^{-}} + \bar\theta^{+}\frac{\partial}{\partial \bar\theta^{-}}\,.\label{cov-1}
\eea
The factors $e^{\pm \frac{i}{2}\mu t_{(A)}}$ appear in \p{cov-1} due to the property that the elements of the supercosets \p{coset-1}, \p{coset-2}
are related to an element of \p{HB} (and to that of the flat ${\cal N}=4, d=1$ harmonic supercoset in the case \p{coset-2}) through multiplication of the latter element
from the right by the exponential $\exp{(-i\mu F)}$ (cf. analogous relations in the non-harmonic case \cite{ISTconf}).

To define the ${\bf (4,4,0)}$ superfields adapted to the supercosets \eqref{coset-1} and \eqref{coset-2}, one should
use the constraints \eqref{q-constr}, \eqref{DY} in which the corresponding covariant derivatives are replaced
by the expressions \p{cov-1}. Note that at any $\alpha \neq 0$ the constraints \eqref{CRsuperf}  give rise to \eqref{A-cond}.
At $\alpha=0$\,, eq. \p{A-cond} should be imposed independently.

The superfield $\epsilon, \bar\epsilon$ transformation law generalizing eq. \p{SFtr} to the harmonic supercosets
\p{coset-1} and \p{coset-2} is as follows (in the original coordinates)
\bea
    \delta \Phi = \mu\left[2\left(1+\alpha\right)\left(\bar{\theta}^{+}\epsilon^{-} - \theta^{+}\bar{\epsilon}^{-}\right) \tilde{F}
     + \alpha\left(\bar{\theta}^{+}\epsilon^{-} + \theta^{+}\bar{\epsilon}^{-}\right) {\cal D}^{0}
    +\alpha\left(\bar{\theta}^{-}\epsilon^{-}+\theta^{-} \bar{\epsilon}^{-} \right){\cal D}^{++}\right]\Phi.
\eea

\subsection{Superconformal Lagrangians for $q^{+a}$}
Skipping details, one can show that all component results for the $SU(2|1)$ superfields $q^{\pm a}$ defined
on the supercosets \eqref{coset-1}, \eqref{coset-2} can be obtained from those for $q^{\pm a}$ on the supercoset \eqref{HB}
by substituting $m = -\alpha \mu$ and redefining the fermionic fields as
\bea
     \psi^a \rightarrow  \psi^a e^{\frac{i}{2}\mu t} ,\qquad\bar{\psi}^a \rightarrow \bar{\psi}^a e^{-\frac{i}{2}\mu t}.\label{redef}
\eea
This redefinition ensures that the $U(1)$ generator $F$ acts only on the fermionic fields and ${\cal H}$ is reduced
to the pure time derivative $i\partial_{t}$ on all fields.

For any $\alpha$, the transformations \eqref{Hypertr} are modified as
\bea
    &&\delta x^{ia} =-\,\epsilon^i\psi^a e^{\frac{i}{2}\mu t}-\bar{\epsilon}^i \bar{\psi}^a e^{-\frac{i}{2}\mu t},\nn
    &&\delta \bar{\psi}_a = \left(2i\epsilon_k \dot{x}^{k}_{a} + \alpha\mu\,\epsilon_k x^{k}_{a}\right) e^{\frac{i}{2}\mu t},\qquad
    \delta \psi^a = \left(2i\bar{\epsilon}^k \dot{x}^{a}_k - \alpha\mu\,\bar{\epsilon}^k x^{a}_k\right)e^{-\frac{i}{2}\mu t}.\label{Hypertrmu}
\eea
The transformations with $-\mu$ amount to the $-\mu$ version of the superalgebra \eqref{alpha}.
The two sets  of transformations  produce as their closure the superconformal group $D(2,1;\alpha)$.

At $\alpha = 0$ the superfields $q^{\pm a}$ live on the supercoset \p{coset-2} corresponding to the $U(1)$-deformed Poincar\'e superalgebra \p{alpha1}.
The relevant component
field transformations are the $\alpha = 0$ version of \eqref{Hypertrmu}. Together with their $-\mu$ counterparts (forming the $-\mu$ analog of the algebra \p{alpha1})
they close on
\bea
    D\left(2,1;0\right)\cong PSU\left(1,1|2\right)\rtimes SU(2).\label{product}
\eea

As was shown in \cite{ISTconf}, the trigonometric superconformal models for the multiplets ${\bf (1,4,3)}$ and ${\bf (2,4,2)}$
possess the notable common property  that their superconformal Lagrangians are functions of $\mu^2$ and so
they are invariant under the reflection $\mu \rightarrow -\mu$. The same feature proves to be inherent to
the supermultiplets ${\bf (4,4,0)}$ too.

After making the substitution \eqref{redef} in the Lagrangian \eqref{LagrKincomp},
the latter proves to contain no terms linear in $\mu$ and so is superconformal for the choice $G =  \left(x^{ia}x_{ia}\right)^{\frac{1-\alpha}{\alpha}}$ only.
The relevant trigonometric superconformal Lagrangian is a deformation of the parabolic superconformal Lagrangian by the oscillator term \cite{HT}
\bea
    {\cal L}_{\rm sc}^{(\alpha)}&=& \bigg[\,\dot{x}^{ia}\dot{x}_{ia}
    +\frac{i}{2}\left(\bar{\psi}_a\dot{\psi}^a-\dot{\bar{\psi}}_a\psi^a\right)
    -\frac{i}{2}\left(\psi_a\bar{\psi}^c+\psi^c\bar{\psi}_a\right)\dot{x}^{ia}\partial_{ic}
    -\frac{1}{16}\left(\bar\psi\,\right)^2\left(\psi\right)^2\Delta_x \nn
     &&-\,\frac{\alpha^2\mu^2}{4}\,x^{ia}x_{ia}\,\bigg]\left(x^{ia}x_{ia}\right)^{\frac{1-\alpha}{\alpha}}.\label{qconf}
\eea
The free Lagrangian corresponds to the choice $\alpha=1$.

In the particular case $\alpha=0$ the expression \p{qconf} becomes singular and, in order to construct the meaningful superconformal action,
one is led to redefine the field $x^{ia}$ as
\bea
    x^{ia}\rightarrow x^{ia} + \frac{\rho^{ia}}{\alpha}\,,
\eea
and then send $\alpha\rightarrow 0$. This gives rise to the inhomogeneous $\rho$ dependent transformations for the $\alpha=0$ case
\bea
    &&\delta x^{ia} =-\,\epsilon^i\psi^a e^{\frac{i}{2}\mu t}-\bar{\epsilon}^i \bar{\psi}^a e^{-\frac{i}{2}\mu t},\nn
    &&\delta \bar{\psi}_a = \left(2i\epsilon_k \dot{x}^{k}_{a} + \mu\,\epsilon_k\rho^{k}_{a} \right) e^{\frac{i}{2}\mu t},\qquad
    \delta \psi^a = \left(2i\bar{\epsilon}^k \dot{x}^{a}_k - \mu\,\bar{\epsilon}^k\rho^{a}_{k}\right)e^{-\frac{i}{2}\mu t}.\label{trq-rho}
\eea
The superconformal Lagrangian for $\alpha=0$ is obtained as the limit
\bea
    {\cal L}_{\rm sc}^{(\alpha=0)}= \lim_{\alpha\rightarrow 0}{\cal L}_{\rm sc}^{(\alpha , \rho)}\,,
\eea
with detaching some singular overall factor in the end. The Lagrangian is given by
\bea
    {\cal L}_{\rm sc}^{(\alpha=0)}&=& \bigg[\,\dot{x}^{ia}\dot{x}_{ia}
    +\frac{i}{2}\left(\bar{\psi}_a\dot{\psi}^a-\dot{\bar{\psi}}_a\psi^a\right)
    -\frac{i}{2}\,\dot{x}^{ia}\left(\psi_a\bar{\psi}^c+\psi^c\bar{\psi}_a\right)\partial_{ic}
    -\frac{1}{16}\left(\bar\psi\,\right)^2\left(\psi\right)^2\Delta_x\nn
    &&-\,\frac{\mu^2\rho^2}{4}\,\bigg]\exp{\bigg\lbrace\frac{2\rho^{ia}x_{ia}}{\rho^2}\bigg\rbrace}.\label{qconf0}
\eea
Here, both the Pauli-G\"ursey symmetry $SU(2)_{\rm PG}$ acting on the indices $a$ and the automorphism $SU(2)$ symmetry with generators $I^i_k$
are broken down to the diagonal subgroup $SU(2)_{\rm diag}$. The internal subgroup $SU^{\prime}(2)_{\rm int} \subset PSU\left(1,1|2\right)$
acts only on the fermionic fields.
So, the Lagrangian \eqref{qconf0} is invariant under the superconformal symmetry $PSU\left(1,1|2\right)\rtimes SU(2)_{\rm diag}\,$.
The parabolic analog of this Lagrangian was considered in \cite{DeldIv2}.

Note that the WZ Lagrangian \eqref{WZq} is $\alpha=0$ superconformal, because it is invariant under both the transformations
\eqref{Hypertrmu} with $\alpha=0$ and their $-\mu$ counterparts. Indeed, it is invariant under the redefinition \p{redef} and the second term
in \p{VarWZ1} disappears, when $m = -\alpha \mu$ and $\alpha =0$. No additional parameters $\rho$ are required in this case.

\subsection{Superconformal Lagrangians for $Y^{+A}$}
Next, we consider the mirror multiplet ${\bf (4,4,0)}$, proceeding from the superspaces \eqref{coset-1}, \eqref{coset-2}.

Passing to the superspaces \eqref{coset-1}, \eqref{coset-2} should be accompanied by the following field redefinitions (cf. \p{redef}):
\bea
    &&Y^{A} \rightarrow Y^A e^{-\frac{i}{2}\mu t_{(A)}} ,\qquad
    \bar{Y}^A \rightarrow\bar{Y}^A e^{\frac{i}{2}\mu t_{(A)}},\nn
    &&y^{A} \rightarrow y^A e^{-\frac{i}{2}\mu t} ,\qquad
    \bar{y}^A \rightarrow\bar{y}^A e^{\frac{i}{2}\mu t},\qquad m=-\,\alpha\mu\,.\label{redef2}
\eea
The constraints \eqref{DY} are now imposed with the derivatives \eqref{cov-1}. The constraints \eqref{FY} become
\bea
     \left(1+\alpha\right)\mu\tilde{F}\,\bar{Y}^{A} = -\,\frac{\left(1+\alpha\right)}{2}\,\mu\,\bar{Y}^{A},
     \qquad \left(1+\alpha\right)\mu\tilde{F}\,Y^{A} =\frac{\left(1+\alpha\right)}{2}\,\mu\,Y^{A}.
\eea

The redefinitions \p{redef2} bring the transformations \eqref{trY} to the form:
\bea
    &&\delta y^{A} =-\,\epsilon_i\psi^{iA} e^{\frac{i}{2}\mu t} ,\qquad
    \delta \bar{y}^A =-\,\bar{\epsilon}_i \psi^{iA}e^{-\frac{i}{2}\mu t},\nn
    &&\delta \psi^{iA} =\bar{\epsilon}^i \left[2i\dot{y}^{A}+\left(1+\alpha\right)\mu\,y^A\right] e^{-\frac{i}{2}\mu t}
    - \epsilon^i\left[2i\dot{\bar{y}}^{A}-\left(1+\alpha\right)\mu\,\bar{y}^A\right] e^{\frac{i}{2}\mu t}.\label{trY-1}
\eea
These transformations, together with those in which the replacement $\mu\rightarrow - \mu$ is made,  generate superconformal $D(2,1;\alpha)$ transformations.

Requiring the superconformal invariance of the $\sigma$-model Lagrangian  obtained from \eqref{kin-Y} through the changes \eqref{redef2} (i.e.
just demanding the terms $\sim \mu$ to cancel), we find that
\bea
    \tilde{G} = \left(y^{A}\bar{y}_{A}\right)^{-\frac{2+\alpha}{1+\alpha}}.
\eea
The resulting component Lagrangian reads
\bea
    \label{LYconf}
    \tilde{\cal L}_{\rm sc}^{(\alpha)} &=& \bigg[2\,\dot{y}^{A}\dot{\bar{y}}_{A} +\frac{i}{2}\, \psi^{iA}\dot{\psi}_{iA}
    -\frac{i}{2}\,\psi^{iA}\psi_{iC}\left(\dot{y}^{C}\partial_A +\dot{\bar y}^{C}\bar{\partial}_{A}\right)
    +\frac{1}{48}\,\psi^{iA}\psi^{k}_{A}\psi_{i}^{B}\psi_{kB}\,\Delta_y\nn
    &&- \,\frac{\left(1+\alpha\right)^2\mu^2}{2}\,y^{A}\bar{y}_{A}\,\bigg]\left(y^{A}\bar{y}_{A}\right)^{-\frac{2+\alpha}{1+\alpha}} .
\eea
It is equivalent to the Lagrangian \eqref{qconf} up to the substitution  $\alpha\rightarrow-\left(1+\alpha\right)$.
The free system with $G=\tilde{G} =1$ corresponds to the choice $\alpha=-2$ in \p{LYconf}.

At $\alpha=-1$ the Lagrangian \p{LYconf} is singular. To construct the superconformal $\sigma$-model term in this case,
one introduces an arbitrary inhomogeneity parameter $\rho^A$ as
\bea
    y^{A}\rightarrow y^{A} + \frac{\rho^{A}}{1+\alpha}\,,\qquad \bar{y}^{A}\rightarrow \bar{y}^{A}
    + \frac{\bar{\rho}^{A}}{1+\alpha}\,,\qquad \rho^2=\rho^A\bar{\rho}_A \label{rhoA}
\eea
and then sends $\alpha\rightarrow -1$. At $\alpha=-1$, the transformations \eqref{trY-1} become
\bea
    &&\delta y^{A} =-\,\epsilon_i\psi^{iA} e^{\frac{i}{2}\mu t} ,\qquad
    \delta \bar{y}^A =-\,\bar{\epsilon}_i \psi^{iA}e^{-\frac{i}{2}\mu t},\nn
    &&\delta \psi^{iA} =\bar{\epsilon}^i \left[2i\dot{y}^{A} + \mu\,\rho^A\right] e^{-\frac{i}{2}\mu t}
    - \epsilon^i\left[2i\dot{\bar{y}}^{A}-\mu\,\bar{\rho}^A\right] e^{\frac{i}{2}\mu t}.\label{trY-rho}
\eea
Respectively, the $\alpha=-1$ superconformal $\sigma$-model Lagrangian is written as  (up to some divergent overall factor which can be thrown out)
\bea
    \label{LYconfrho}
    \tilde{\cal L}_{\rm sc}^{(\alpha = -1)} &=& \bigg[2\,\dot{y}^{A}\dot{\bar{y}}_{A} +\frac{i}{2}\, \psi^{iA}\dot{\psi}_{iA}
    -\frac{i}{2}\,\psi^{iA}\psi_{iC}\left(\dot{y}^{C}\partial_A +\dot{\bar y}^{C}\bar{\partial}_{A}\right)
    +\frac{1}{48}\,\psi^{iA}\psi^{k}_{A}\psi_{i}^{B}\psi_{kB}\,\Delta_y\nn
    &&-\,\frac{\mu^2\rho^2}{4}\,\bigg]\exp{\left\lbrace -\frac{y^A\bar{\rho}_A+\rho^A\bar{y}_A}{\rho^2}\right\rbrace} .
\eea
This Lagrangian can be brought into the same form as the previous $\alpha=0$ Lagrangian \eqref{qconf0} by a simple relabeling of the involved fields.
The superconformal group is reduced to $D\left(2,1;-1\right)\cong PSU\left(1,1|2\right)\rtimes SU^\prime(2)_{\rm diag}$,
where $SU^{\prime}(2)_{\rm diag}$ is the diagonal subgroup in the product of the external  $SU^\prime(2)$
(to which the decoupled generator $F$ belongs) and $SU^\prime (2)_{\rm PG}$ (which acts on the indices $A$).

The redefinitions \p{redef2} bring the WZ Lagrangian \p{WZ-Y} to the form
\bea
\tilde{\cal L}_{\rm WZ}^{(\alpha)}= 2\gamma\, \Big\lbrace i \left(\dot{y}^{A}\,\partial_{A}f - \dot{\bar{y}}^{A}\,\bar{\partial}_{A}f\right)
     - (1+\alpha)\frac{\mu}{2}\left({y}^{A}\,\partial_{A}f + {\bar{y}}^{A}\,\bar{\partial}_{A}f\right)
     - \frac{1}{2}\,\psi^{iA}\psi^{B}_{i}\,\partial_{A}\bar{\partial}_{B}f\,\Big\rbrace.\label{WZ-Yalpha}
\eea
We observe that the trigonometric superconformal WZ Lagrangian can be defined only at $\alpha=-1\,$, when the term $\sim \mu$ drops out.
This Lagrangian is written as
\bea
    \tilde{\cal L}_{\rm scWZ}^{(\alpha=-1)} = 2\gamma\, \Big\lbrace i \left(\dot{y}^{A}\,\partial_{A}f - \dot{\bar{y}}^{A}\,\bar{\partial}_{A}f\right)
     - \frac{1}{2}\,\psi^{iA}\psi^{B}_{i}\,\partial_{A}\bar{\partial}_{B}f\,\Big\rbrace, \label{WZ-Yconf}
\eea
and it is invariant (up to a total derivative)  under the homogeneous transformations \eqref{trY-1} with $\alpha = -1$ and their $\mu \rightarrow -\mu$ counterparts.
Thus the corresponding action reveals invariance under their closure $PSU(1,1|2)$.

The function $f$ satisfies the self-duality conditions \eqref{self}, while the condition \eqref{U11} is modified as
\bea
    \left(1+\alpha\right)\mu\,\tilde{F}f\left(Y,\bar{Y}\right) = 0\qquad\Rightarrow\qquad \left(1+\alpha\right)\mu
    \left(y^B\partial_B -\bar{y}^B\bar{\partial}_{B}\right)f\left(y,\bar{y}\right)  =  0\,.
\eea
At $\alpha=-1$ this condition is automatically satisfied and so does not impose any restrictions on the function $f$ in \eqref{WZ-Yconf}.

As we briefly discussed in the previous subsection, WZ term \eqref{WZq} is superconformal at $\alpha=0$\,.
Using the definition \eqref{4v-A}, it is easy to show the equivalence of \eqref{WZ-Yconf} and \eqref{WZq}, as well as the equivalence
of the relevant superconformal transformations\footnote{Note that this equivalence gets broken if we simultaneously consider WZ terms
for both types of the ${\bf (4,4,0)}$ multiplet. This sum cannot be made superconformal by any choice of $\alpha$.}. In
the next subsection we consider infinite-dimensional symmetries inherent to these WZ terms. Note that the sum of the Lagrangians \p{WZ-Yconf} and
\p{LYconfrho} {\it is not} superconformal, since the supergroup $PSU(1,1|2)$ is realized in these Lagrangians by transformations
of the different kinds (these are inhomogeneous for \p{LYconfrho} and homogeneous for \p{WZ-Yconf}).

\subsection{The centerless ${\cal N}=4$ super Virasoro algebra}\label{sVir}
The superconformal WZ Lagrangian \eqref{WZ-Yconf} is not deformed by $\mu$\,.
Hence, it is simultaneously invariant under the following transformations\footnote{They coincide with the $\mu=0$ case of the transformations \eqref{trY-1}\,.}
\bea
    \delta y^{A} =-\,\eta_i\psi^{iA},\qquad
    \delta \bar{y}^A =-\,\bar{\eta}_i \psi^{iA},\qquad
    \delta \psi^{iA} = 2i\bar{\eta}^i\dot{y}^{A}
    - 2i\eta^i\dot{\bar{y}}^{A}\,,
\eea
which generate the standard ``Poincar\'e'' ${\cal N}=4$, $d=1$ supersymmetry.
The Lie brackets of these additional transformations with both \eqref{trY-1} and the $-\mu$ counterparts of \eqref{trY-1} produce the new bosonic generators
$$\sim e^{\pm\frac{i}{2}\mu t}\,i\partial_t\,.$$
On the other hand, generators of the conformal algebra $so(2,1)\subset psu(1,1|2)$ have the following trigonometric realization \cite{ISTconf}:
\bea
    {\cal H}=i\partial_t\,,\qquad T = e^{-i\mu t}\,i\partial_t\,,\qquad\bar{T} = e^{i\mu t}\,i\partial_t\,.
\eea
Commuting the new bosonic generators with these ones, we find that the bosonic subalgebra extends to an infinite-dimensional Virasoro algebra \cite{HoTo}
\bea
    \left[L_k, L_n\right] = \left(k-n\right)L_{k+n}\,,\qquad
    L_n= \frac{2i}{\mu}\,e^{\frac{i}{2} n\mu t}\,\partial_t\,,\qquad k,n\in\mathbb{Z}\,.
\eea
Computing further their commutators with supercharges, we finally find, as the full symmetry of \p{WZ-Yconf}, the Ramond version
of the centerless (small) ${\cal N}=4$ super Virasoro algebra \cite{HT}.

The isomorphic super Virasoro algebra extending the finite-dimensional $\alpha=0$ superconformal algebra $psu(1,1|2)$ is a  symmetry
of the WZ component Lagrangian \eqref{WZq}.

As was noticed in  \cite{HT}, in regard to the ${\bf(4,4,0)}$ multiplet the centerless super Virasoro group can possess
only homogeneous ($\rho = 0$) realizations with the scaling dimension $\lambda_D = 0\,$. This is consistent with the absence
of the superconformal WZ action for $q^{\pm a}$ at $\alpha\neq 0$,
since $\alpha$ is identified as $\alpha= -\,2\lambda_D$ \cite{kuto,khto,HT}. The same conclusion is  valid for
the superfields $Y^A, \bar{Y}^A$, with $\alpha= 2\lambda_D - 1$\,.

\section{Concluding remarks and outlook}
In this paper we generalized the $d=1$ harmonic superspace approach \cite{IvLe} to the case of worldline realizations of the supergroup $SU(2|1)$ as a deformation
of the flat ${\cal N}=4, d=1$ supersymmetry and constructed the general superfield and component actions for the $SU(2|1)$ analogs of the ``root'' ${\cal N}=4, d=1$
multiplet ${\bf (4, 4, 0)}$ and its ``mirror'' version. We also selected the superconformal subclass of general $SU(2|1)$ invariant actions of these multiplets.

In contrast to the flat harmonic superspace, there is no direct equivalence between the two types of the ${\bf (4,4,0)}$ supermultiplet.
In the case of the standard supermultiplet ${\bf (4,4,0)}$, the $SU(2)_{\rm int}$ subgroup of $SU(2|1)$ is realized only
on the bosonic fields $x^{ia}$ of $q^{\pm a}$. For the mirror multiplet ${\bf (4,4,0)}$, this group is realized on the fermionic fields $\psi^{iA}$ of $Y^{A}, \bar{Y}^A$.
One of the manifestations of the non-equivalence just mentioned is the non-existence of the $SU(2|1)$ invariant WZ action for $q^{\pm a}$
and the existence of such an action for the mirror multiplet.
On the other hand, the trigonometric (as well as the parabolic) superconformal models of both ${\bf (4,4,0)}$ multiplets are equivalent to each other,
up to the substitutions
\bea
    x^{ia}\leftrightarrow y^{i^\prime A},\qquad \psi^{i^\prime a}\leftrightarrow \psi^{i A} \qquad\alpha \leftrightarrow -\left(1+\alpha\right).
\eea
As is known, the interchange $\alpha \leftrightarrow -\left(1+\alpha\right)$ amounts to permuting the $SU(2)$ and $SU^{\prime}(2)$
generators in $D\left(2,1;\alpha\right)$ and so is an automorphism of this superalgebra.

The scaling dimension parameter $\lambda_D$ is identified with $\alpha = -\,2\lambda_D$ for the standard multiplet ${\bf (4,4,0)}$ and with $\alpha = 2\lambda_D -1$
for the mirror multiplet ${\bf (4,4,0)}$.
The superconformal group $D\left(2,1;\alpha\right)$ with $\lambda_D = 0$ can be reduced to the $PSU(1,1|2)\rtimes SU(2)$ supergroup for both supermultiplets,
at $\alpha=0$ for $q^{\pm a}$ and $\alpha = -1$ for $Y^A, \bar{Y}^A$.

For reader's convenience, in the Table \ref{table1} we summarize the results concerning symmetries of the $\sigma$-model and WZ actions.
An important role is played by the inhomogeneity parameters $\rho$ allowing to construct superconformal $\sigma$-model actions for the special case $\lambda_D=0$\,,
so that they are invariant under the $\rho$-modified inhomogeneous transformations \eqref{trq-rho}, \eqref{trY-rho}.
On the other hand, superconformal WZ actions at $\lambda_D=0$ are invariant under the homogeneous superconformal transformations given by \eqref{Hypertrmu} and \eqref{trY-1}
(with $\alpha =0$ and $\alpha=-1$, respectively).

In the case of the mirror multiplet, we can consider $SU(2|1)$ $\sigma$-model actions combined with WZ actions
(the Lagrangian \eqref{LYfree} provides an example of such a system).
The combined actions, being $SU(2|1)$ invariant,  cannot be made superconformal.

A few comments are in order concerning the third column in the Table \ref{table1}. For the non-conformal models,
the descriptions in the standard flat ${\cal N}=4, d=1$ harmonic superspace \cite{IvLe}
and its $U(1)$-deformed analog \p{coset-2} are completely equivalent. The only seeming difference
is that in the case of \p{coset-2} the supersymmetry transformations produce not only the standard shift
of $t_{(A)}$ (associated as a coset coordinate with the generator ${\cal H}$) but also some induced
$U(1)$ transformation with the stability subgroup generator $F$. However, this difference can be
leveled just by redefining all the $U(1)$ charged fields on the pattern of redefinitions \p{redef} or \p{redef2}.
After that the canonical time-translation generator becomes ${\cal H} + \mu F$ in \p{alpha1} and
one recovers the standard ${\cal N}=4, d=1$ picture. On the other hand, in the superconformal case
it is the description in \p{coset-2} that allows one to define two sets of the
$\mu$-deformed ${\cal N}=4, d=1$ supersymmetry transformations related via the change $\mu \rightarrow -\mu$
and to construct the superconformal models with the trigonometrical realization
of the supergroup $D(2,1;0)$ as a closure of these two sets. In the standard undeformed
${\cal N}=4, d=1$ harmonic superspace only parabolic realizations of superconformal symmetries are achievable.\\

\begin{table}[ht]
\begin{center}
\begin{footnotesize}
\begin{tabular}{|l|c|c|c|c|}
\hline
                                                & ${\cal N}=4$, $d=1$   & $SU(2|1)$ & $U(1)$-deformed   & Superconformal                                \\
\hline
        $\sigma$-model actions for $q^{\pm a}$     & $+$                   & $+$       & $+$               & ($\alpha\neq0$) or ($\alpha=0$\,, $\rho$) \\
\hline
        WZ actions for $q^{\pm a}$              & $+$                   & $-$       & $+$               & ($\alpha=0$)                                  \\
\hline
        $\sigma$-model $+$ WZ for $q^{\pm a}$          & $+$                   & $-$       & $+$               & $-$                                           \\
\hline
        $\sigma$-model actions for $Y^{i^\prime a}$    & $+$                   & $+$       & $+$               & ($\alpha\neq-1$) or ($\alpha=-1$\,, $\rho$)   \\
\hline
        WZ actions for $Y^{i^\prime a}$         & $+$                   & $+$       & $+$               & ($\alpha=-1$)                                 \\
\hline
        $\sigma$-model $+$ WZ for $Y^{i^\prime a}$ & $+$                   & $+$       & $+$               & $-$                                           \\
\hline
\end{tabular}
\captionof{table}{First 2 columns show existence of $\sigma$-model and WZ actions based on the flat harmonic superspace and the $SU(2|1)$ harmonic superspace.
The third column corresponds to the harmonic superspace \eqref{coset-2}.
The last column marks the parameters of superconformal $\sigma$-model and WZ actions of the parabolic and trigonometric types.}\label{table1}
\end{footnotesize}
\end{center}
\end{table}

As for further developments, here we would like to mention the construction
of SQM models for the $SU(2|1)$ analog of the multiplet ${\bf (3, 4, 1)}$ which in the flat case has a natural description in the
analytic harmonic superspace, the construction of the multi-particle SQM models with various types of the $SU(2|1)$ multiplets taken
into account, and generalizations of the harmonic superspace approach to higher rank deformed $d=1$ supersymmetries, e.g., $SU(2|2)$ which
can be regarded as a deformation of the flat ${\cal N}=8, d=1$ supersymmetry \cite{DSQM}. Besides this, it seems of interest
to explicitly solve some more complicated examples of the SQM $SU(2|1)$ models, not only the simplest ones treated here and in \cite{DSQM}.
Another problem to be clarified is whether the $SU(2|1)$ analogs of the nonlinear ${\bf (4, 4, 0)}$ multiplets \cite{DeldIv4} exist and
what could be the $m$ deformation of the target geometries associated with such multiplets.
It would be also interesting to define and study $SU(2|1)$
analogs of various useful concepts of the flat ${\cal N}=4, d=1$ supersymmetry, such as the semi-dynamical spin multiplets \cite{FeIvLe}, the gauging procedure
in the ${\cal N}=4$ SQM models \cite{DeldIv1,DeldIv2,DeldIv3}, etc. All these methods and concepts essentially use the notions of the $d=1$
harmonic superspace approach. There also remains the problem of recovering various $SU(2|1)$ SQM models through the direct dimensional reduction from the
higher-dimensional theories with the curved analogs of the Poincar\'e supersymmetry. Recently, a variant of the deformed SQM was applied to compute the vacuum (Casimir)
energy in some conformal field theories \cite{Casimir}. It is interesting to establish possible links of this construction with the deformations
of the ${\cal N}=4, d=1$ supersymmetry (and its  ${\cal N}=2$ reductions) considered in \cite{DSQM} - \cite{ISTconf} and in the present paper.

\section*{Acknowledgements}
We are grateful to Armen Nersessian and Andrei Smilga for interest in various aspects of $SU(2|1)$ SQM models, useful suggestions
and comments. This work was partially supported by the RFBR grant Nr. 15-02-06670 and a grant of the Heisenberg-Landau Program.

\appendix

\section{Transformations of the covariant derivatives}
Here we collect the $\epsilon, \bar\epsilon$ transformations of the covariant derivatives constituting the CR-structure \p{CRstruct}, as well
as of the derivative ${\cal D}^{--}$:
\bea
\delta{\cal D}^{++} &=& m\left(1-m\,\bar{\theta}^{+}\theta^{-}+m\,\bar{\theta}^{-}\theta^{+}\right)
\left(\bar{\theta}^{+}\epsilon^{+}+\theta^{+} \bar{\epsilon}^{+} \right){\cal D}^{0}
    -m\left(\bar{\theta}^{+}\epsilon^{-}+\theta^{+} \bar{\epsilon}^{-} \right){\cal D}^{++}\nn
    &&+\,m\left(1-m\,\bar{\theta}^{+}\theta^{-}+m\,\bar{\theta}^{-}\theta^{+}\right)\left(\bar{\theta}^{-}\epsilon^{+}+\theta^{-} \bar{\epsilon}^{+} \right){\cal D}^{++}
    \nn
    && + \,2 m\left(1-m\,\bar{\theta}^{+}\theta^{-}+m\,\bar{\theta}^{-}\theta^{+}\right)
    \left(\bar{\theta}^{+}\epsilon^{+}-\theta^{+} \bar{\epsilon}^{+} \right)\tilde{F},\nn
\delta{\cal D}^{+} &=& -2m\,\theta^{+}\bar{\epsilon}^{-}{\cal D}^{+} + m\left[\bar{\epsilon}^{-}
+ m\,\bar{\theta}^{-}\left(\theta^{-}\bar{\epsilon}^{+}+\bar{\theta}^{+}\epsilon^{-}+\theta^{+} \bar{\epsilon}^{-} \right)\right]{\cal D}^{++}
    \nn
    &&+\,m^2\,\bar{\theta}^{-}\left(1-m\,\bar{\theta}^{+}\theta^{-}\right)\left[\left(\bar{\theta}^{+}\epsilon^{+}+\theta^{+} \bar{\epsilon}^{+} \right){\cal D}^{0}
    +2\left(\bar{\theta}^{+}\epsilon^{+}-\theta^{+} \bar{\epsilon}^{+} \right)\tilde{F}\right],\nn
    \delta\bar{{\cal D}}^{+} &=& -2m\,\bar{\theta}^{+}\epsilon^{-}\bar{{\cal D}}^{+}
    -\, m\left[\epsilon^{-}+m\,\theta^{-}\left(\bar{\theta}^{-}\epsilon^{+}+\bar{\theta}^{+}\epsilon^{-}+\theta^{+} \bar{\epsilon}^{-} \right) \right]{\cal D}^{++}
     \nn
    &&-\,m^2\,\theta^{-}\left(1+m\,\bar{\theta}^{-}\theta^{+}\right)\left[\left(\bar{\theta}^{+}\epsilon^{+}+\theta^{+} \bar{\epsilon}^{+} \right){\cal D}^{0}
    +2\left(\bar{\theta}^{+}\epsilon^{+}-\theta^{+} \bar{\epsilon}^{+} \right)\tilde{F}\right], \nn
\delta{\cal D}^{--} &=& 2 m\left(\bar{\theta}^{+}\epsilon^{-}+\theta^{+}\bar{\epsilon}^{-} \right){\cal D}^{--}
+ 2 m\left(\bar{\theta}^{-}\epsilon^{-}-\theta^{-} \bar{\epsilon}^{-} \right)\tilde{F},\qquad \delta{\cal D}^{0} = 0\,. \lb{TransfCovDer}
\eea
These variations are obtained just by making use of the coordinate transformations \p{HBtr}. Note that the redefined harmonic derivative
\bea
D^{++} &:=& \left(1+m\,\bar{\theta}^{+}\theta^{-}-m\,\bar{\theta}^{-}\theta^{+}\right){\cal D}^{++} = \partial^{++}
+ 2i \,\theta^+\bar\theta^+\partial_A - 2 m\,\theta^+\bar\theta^{+}\tilde{F} \nn
&&    +\, \left(1+m\,\bar{\theta}^{+}\theta^{-}-m\,\bar{\theta}^{-}\theta^{+}\right)\left[\theta^{+}\frac{\partial}{\partial \theta^{-}}
+ \bar\theta^{+}\frac{\partial}{\partial \bar\theta^{-}}\right]
\eea
has a simpler transformation law
\be
\delta D^{++} = m\left(\bar{\theta}^{+}\epsilon^{+}+\theta^{+} \bar{\epsilon}^{+} \right){\cal D}^{0}
    + 2 m\left(\bar{\theta}^{+}\epsilon^{+}-\theta^{+} \bar{\epsilon}^{+} \right)\tilde{F}\,.
\ee

Despite the rather involved form of \p{TransfCovDer}, the objects ${\cal D}^{\pm\pm}\Phi, \;{\cal D}^+\Phi$ and $\bar{\cal D}^{+}\Phi$
transform according to the simple universal transformation law \p{SFtr}, e.g.,
$$
\delta ({\cal D}^{++}\Phi) = -m\left[2 \left(\bar{\theta}^{+}\epsilon^{-} - \theta^{+}\bar{\epsilon}^{-}\right) \tilde{F} + \left(\bar{\theta}^{+}\epsilon^{-}
    + \theta^{+}\bar{\epsilon}^{-}\right) {\cal D}^{0}
    +\left(\bar{\theta}^{-}\epsilon^{-}+\theta^{-} \bar{\epsilon}^{-} \right){\cal D}^{++}\right]({\cal D}^{++}\Phi)\,,
$$
i.e. these are $SU(2|1)$ superfields. The rest of covariant derivatives,
i.e. ${\cal D}^-\Phi$ and $\bar{\cal D}^-\Phi$, can be obtained by acting of ${\cal D}^{--}$ on the basis ones forming CR structure,
so they are also transformed by the law \p{SFtr}.
\section{Relation to the central basis of $SU(2|1)$ superspace}
The $SU(2|1)$ superspace in the central basis amounts to the coordinate set \cite{DSQM, SKO, ISTconf}
\bea
    \zeta=\left(t,\theta_{i}, \bar{\theta}^{j}\right). \label{dB}
\eea
Extending these coordinates by the harmonic coordinates $w^{\pm}_i$, we arrive at the central basis of the harmonic $SU(2|1)$ superspace:
\bea
    \zeta_C=\left(t,\theta_{i}, \bar{\theta}^{j},w^{\pm}_i\right).\label{CB}
\eea
The odd $SU(2|1)$  transformations of these coordinates are
\bea
    &&\delta\theta_{i}=\epsilon_{i}+2m\,\bar{\epsilon}^k\,\theta_k \,\theta_{i}\,,\qquad\delta\bar{\theta}^{j}
    =\bar{\epsilon}^{i}-2m\,\epsilon_k\,\bar{\theta}^k\,\bar{\theta}^{i},\qquad\delta t=i\left(\epsilon_k\,\bar{\theta}^k+\bar{\epsilon}^k\,\theta_k  \right),\nn
    &&\delta w^+_{i}= \lambda^{++} w^-_{i},\quad \lambda^{++}=
    -\,m\left(1 - m\,\bar{\theta}^{l}\theta_{l}\right)\left(\bar{\theta}^{k}\epsilon^{j}+\theta^{k} \bar{\epsilon}^{j} \right)w^+_{k}w^+_{j},
    \qquad \delta w^-_{i}=0\,.\label{CBtr}
\eea
The relation with the analytic basis coordinates \p{HB} is given by
\bea
    &&\theta^{i}w^{-}_{i}=\theta^{-}  ,\qquad \theta^{i} w^{+}_{i}=
    \theta^{+} \left(1 + m\,\bar{\theta}^{+}\theta^{-}- m\,\bar{\theta}^{-}\theta^{+}\right),\nn
    &&\bar{\theta}^{k} w^{-}_k = \bar{\theta}^{-},\qquad \bar{\theta}^{k} w^{+}_k=\bar{\theta}^{+}
    \left(1 + m\, \bar{\theta}^{+}\theta^{-}- m\,\bar{\theta}^{-}\theta^{+}\right),\nn
    && t = t_{(A)} + i\left(\bar{\theta}^{-}\theta^{+} + \bar{\theta}^{+}\theta^{-}\right).\label{subs}
\eea
It is direct to show that the transformations \p{CBtr} in the central basis yield for the analytic basis coordinates just
the transformations \eqref{HBtr}.

Using the above correspondence, the supermultiplets $({\bf 4,4,0})$  can also be described in the central basis.
The corresponding superfield $q^{ia}$ does not depend on harmonics and obeys the $SU(2|1)$ covariant constraints
\bea
    {\cal D}^{(k}q^{i)a} = \bar{{\cal D}}^{(k}q^{i)a} =0\,, \qquad \tilde{F}q^{ia} = 0\,, \qquad \left( q^{ia}\right)^\dagger =q_{ia}\,.
\eea
The expressions for ${\cal D}^{k}, \bar{{\cal D}}^{k}$ were given in \cite{DSQM}.
Solving these constraints, we find
\bea
    q^{ia}\left(\zeta\right)&=&\left[1+ \frac{m}{2}\,\bar{\theta}^k\theta_k - \frac{5m^2}{16}\left(\bar{\theta}\,\right)^2\left(\theta\right)^2\right]x^{ia}
    +\left(1-\frac{m}{2} \,\bar{\theta}^k\theta_k\right)\left(\theta^i \psi^a
    +\bar{\theta}^i\bar{\psi}^a \right)\nn
    && - \,i \epsilon_{kl}\left(\bar{\theta}^i \theta^l + \bar{\theta}^l \theta^i\right)\dot{x}^{ka}
    -i\,\bar{\theta}^k\theta_k\left(\theta^i \dot{\psi}^a
    -\bar{\theta}^i\dot{\bar{\psi}}^a\right) +\frac{1}{4}\left(\bar{\theta}\,\right)^2\left(\theta\right)^2\ddot{x}^{ia}.\label{q}
\eea
In contrast to the harmonic analytic superfield \eqref{q+}, this superfield explicitly involves the deformation parameter $m$.
The superfields $q^{\pm}_{a}$ and $q^{ia}$ are related to each other in the following way:
\bea
    q^{ia}w^{+}_i =q^{+a}\left(1 + m\,\bar{\theta}^{+}\theta^{-}- m\,\bar{\theta}^{-}\theta^{+}\right)^{\frac{1}{2}},\quad
    q^{ia}w^{-}_i =q^{-a}\left(1 + m\,\bar{\theta}^{+}\theta^{-}- m\,\bar{\theta}^{-}\theta^{+}\right)^{-\frac{1}{2}},
\eea
\bea
    q^{ia} =q^{-a}w^{+i}\left(1 + m\,\bar{\theta}^{+}\theta^{-}- m\,\bar{\theta}^{-}\theta^{+}\right)^{-\frac{1}{2}}-q^{+a}w^{-i}
    \left(1 + m\,\bar{\theta}^{+}\theta^{-}- m\,\bar{\theta}^{-}\theta^{+}\right)^{\frac{1}{2}}.
\eea

One can also rewrite the general $q$ action as an integral over the superspace \p{dB}
\bea
    S(q^{ia})= -\int d\zeta\,L\left(q^{2}\right),\qquad q^{2}=q^{ia}q_{ia}\,,
\eea
where
\bea
    d\zeta = dt\, d^2\theta\, d^2\bar{\theta}\left(1+2m\,\bar{\theta}^k\theta_k\right)\lb{InvMeas}
\eea
is the $SU(2|1)$ invariant integration measure. Note that the analytic basis measure \p{ZetaHtr} is related to \p{InvMeas} as
\be
d\zeta_H = dw d\zeta\,.
\ee
It is not $SU(2|1)$ invariant because the harmonic measure $dw$ is not.

In the  superspace \eqref{dB}, the $SU(2|1)$ constraints \p{DY} defining the mirror ${\bf (4, 4, 0)}$ multiplet
are rewritten as
\bea
    \bar{\cal D}^{i} Y^{A} ={\cal D}^{i} \bar{Y}^{A} = 0\,,
    \quad {\cal D}^{i}Y^{A}=\bar{\cal D}^{i}\bar{Y}^{A}.\label{Y-constr}
\eea
Their solution reads
\bea
    Y^{A}\left(\zeta\right) &=& \left(1+2m\,\bar{\theta}^{k}\theta_{k}\right)^{-\frac{1}{4}}\bigg\lbrace y^{A} + \theta_{i} \psi^{iA}
    + i\, \bar{\theta}^{k}\theta_{k} \,\dot{y}^{A}
    + \theta_{k}\theta^{k}\left( i\dot{\bar{y}}^{A} +\frac{m}{2}\,\bar{y}^{A}\right)+ i\,\bar{\theta}^{k}\theta_{k} \theta_{i}\dot{\psi}^{iA}\nn
    &&-\,\frac{1}{4}\left(\bar{\theta}\,\right)^2\left(\theta\right)^2\left(\ddot{y}^A + 2im\dot{y}^A\right)\bigg\rbrace ,\nn
    \bar{Y}^{A}\left(\zeta\right) &=& \left(1+2m\,\bar{\theta}^{k}\theta_{k}\right)^{-\frac{1}{4}}\bigg\lbrace\bar{y}^{A}
    + \bar{\theta}_{i} \psi^{iA} - i\,\bar{\theta}^{k}\theta_{k} \, \dot{\bar{y}}^{A}
    + \bar{\theta}^{k} \bar{\theta}_{k} \left( i\dot{y}^{A} - \frac{m}{2}\,y^{A}\right)
    - i\,\bar{\theta}^{k}\theta_{k} \bar{\theta}_{i}\dot{\psi}^{iA}\nn
    &&-\,\frac{1}{4}\left(\bar{\theta}\,\right)^2\left(\theta\right)^2 \left(\ddot{\bar{y}}^A -2im\dot{\bar{y}}^A\right)\bigg\rbrace .
\eea
The precise relation with the analytic basis solution  $Y^A\big(\zeta_{(A)}^{(3)}\big)$ given in \p{Y} is
\bea
    Y^{A}\left(\zeta\right) = e^{\frac{i}{2}m(t_{(A)} -t)}Y^{A}\big(\zeta_{(A)}^{(3)}\big).
\eea
The general $Y, \bar Y$ action can also be written as an integral over \p{dB}
\bea
    \tilde{S}\left(Y,\bar{Y}\right) = \int d\zeta \,\tilde{L}\left( Y,\bar{Y}\right).
\eea

Finally, we note the existence of another analytic basis in the $SU(2|1)$ harmonic superspace,
$\tilde{\zeta}_H = \left(t_{(A)}, \tilde{\theta}^{\pm}, \bar{\tilde\theta}^{\pm},w^{\pm}_i\right)\,,$ where
\bea
    &&\tilde{\theta}^{+} =\theta^{+}e^{\frac{i}{2}\mu t_{(A)}},\qquad
    \tilde{\theta}^{-} =\theta^{-}e^{\frac{i}{2}\mu t_{(A)}}\left[1-\left(1+\alpha\right)\mu\,\bar{\theta}^{-}\theta^{+}\right],\qquad m=-\,\alpha\mu\,,\nn
    &&\bar{\tilde{\theta}}^{+} =\bar{\theta}^{+}e^{-\frac{i}{2}\mu t_{(A)}},\qquad
    \bar{\tilde{\theta}}^{-} =\bar{\theta}^{-}e^{-\frac{i}{2}\mu t_{(A)}}\left[1+\left(1+\alpha\right)\mu\,\bar{\theta}^{+}\theta^{-}\right].\label{subs2}
\eea
The $SU(2|1)$ transformations (and their $\alpha =0$ degenerate case) in this basis are easily extended
to the trigonometric superconformal ones by making the reflection $\mu \rightarrow -\mu$ and considering a closure of two such sets
of transformations.

The coordinates $\left(t, \tilde{\theta}_i, \bar{\tilde\theta}^k\right)$ introduced in \cite{ISTconf} and ensuring a similar superconformal closure in the
central basis, are related to \p{subs2} by the following simple redefinitions:
\bea
    \tilde{\theta}^{\pm} = \tilde{\theta}^i w^{\pm}_{i}\,,\qquad
     \bar{\tilde{\theta}}^{\pm} = \bar{\tilde{\theta}}^i w^{\pm}_{i}\,,\qquad  t_{(A)} = t - i\left(\bar{\tilde\theta}^{-}\tilde{\theta}^{+}
     + \bar{\tilde\theta}^{+}\tilde{\theta}^{-}\right).\label{subs1}
\eea
It is straightforward to check that the analytic subspace $\left(t_{(A)}, \bar{\tilde\theta}^{+}, \tilde{\theta}^{+}, w^{\pm}_i\right)$ is closed under both
sets of the $SU(2|1)$ transformations and hence under their superconformal closure. The same is true for the $\alpha =0$ case.

\section{$SU(2|1)$ representations}\label{appB}
The finite-dimensional $SU(2|1)$ representations \cite{repres} are characterized by two parameters $\lambda$ and $\beta$.
The number $\lambda$ (``highest weight'') is positive half-integer or integer,
and an arbitrary additional real number $\beta\,$ is related to the eigenvalues of the $U(1)$ generator $\tilde{H}$ of \eqref{algebra-1}.
Casimir operators of the $su(2|1)$ superalgebra \eqref{algebra-1} are written in the model-independent way as follows:
\bea
    &&m^2C_2 = \tilde{H}^2 - \frac{m^2}{2}\, I^i_j I^j_i + \frac{m}{4}\left[Q^i,\bar{Q}_i\right],\label{C2}\\
    &&m^3 C_3 = \frac{m^2}{2}\left(1+2C_2\right)\tilde{H} - \frac{m}{8}\left[m I^i_k - \delta^i_k\,\tilde{H}\right]\left[Q^k,\bar{Q}_i\right].\label{C3}
\eea
The eigenvalues of these Casimir operators can be written as
\bea
    C_2 =\beta^2-\lambda^2 ,\qquad
    C_3 = \beta\, C_2\,.\label{Casimir}
\eea
Non-zero values of Casimir operators define the typical representations of $SU(2|1)$.
In the case $|\beta|=\lambda$, Casimir operators have zero eigenvalues and the relevant $SU(2|1)$ representations are atypical.
Typical and atypical representations have the dimensions $8\lambda$ and $4\lambda+1\,$, respectively.
The fundamental $SU(2|1)$ representation is atypical and it corresponds to the choice $|\beta|=\lambda=1/2$\,.

\end{document}